\documentclass[superscriptaddress,twocolumn,english,floatfix,aps,prd,amsmath,amssymb,longbibliography]{revtex4-2}
\usepackage[T1]{fontenc}
\usepackage[latin9]{inputenc}
\usepackage{color}
\usepackage{times}
\usepackage{babel}
\usepackage{epsfig}
\usepackage{subfigure}
\usepackage{gensymb}
\usepackage{textcomp}
\usepackage{amsmath}
\usepackage{amssymb}
\usepackage{amsbsy}
\usepackage{graphicx}
\usepackage{float}
\usepackage{esint}
\usepackage[colorlinks=true]{hyperref}
\hypersetup{citecolor=blue,linkcolor=blue,urlcolor=blue}

\begin{document}

\title{Introducing the notion of tensors through a variation of a Feynman didactic approach}

\author{Lucas Queiroz}
\email{lucas.silva@icen.ufpa.br}
\affiliation{Faculdade de F\'{i}sica, Universidade Federal do Par\'{a}, 66075-110, Bel\'{e}m, Par\'{a}, Brazil}

\author{Edson C. M. Nogueira}
\email{edson.moraes.nogueira@icen.ufpa.br}
\affiliation{Faculdade de F\'{i}sica, Universidade Federal do Par\'{a}, 66075-110, Bel\'{e}m, Par\'{a}, Brazil}

\author{Danilo T. Alves}
\email{danilo@ufpa.br}
\affiliation{Faculdade de F\'{i}sica, Universidade Federal do Par\'{a}, 66075-110, Bel\'{e}m, Par\'{a}, Brazil}
\affiliation{Centro de F\'{i}sica, Universidade do Minho, P-4710-057, Braga, Portugal}

\date{\today}

\begin{abstract}
In one of his books [\textit{The Feynmann Lectures on Physics}, vol. 2],
Feynman presents a didactic approach to introduce basic ideas about tensors, using, as a first example, the dependence of the induced polarization of a crystal on the direction of the applied electric field,
and also presenting the energy ellipsoid as a way of visualizing the polarization tensor. 
In the present paper, we propose 
some variations on Feynman's didactic approach,
considering as our basic models a single ground-state atom and a carbon dioxide ($\text{CO}_{2}$) molecule, instead of crystals, and introducing a visual representation of tensors based on the ideas of the Lamé stress ellipsoid, instead of the energy ellipsoid.
With these changes, the resulting didactic proposal presents a reduction in the prerequisites of physical and mathematical concepts if compared to Feynman's original approach, requiring, for example, no differential calculus and only introductory vector algebra.
The text is written so that it can be used directly as a learning tool for 
students (even those in the beginning of the undergraduate course), 
as well as for teachers interested in preparing their own materials.
\end{abstract}
%
\maketitle
%
\section{Introduction}
\label{intro}

A physical system whose properties are the same in all directions is called isotropic, whereas when they are different in distinct directions, it is called anisotropic.
Feynmam, in Chapter 31 of Ref. \cite{Feynman-Lectures-vol-2}, points that students in undergraduate courses have to deal, at some moment of their studies or future careers, with real situations involving anisotropic properties, as the electric conductivity, moment of inertia, stress caused by a force acting on a body, among others. 
According to him, undergraduate students, in courses of basic physics, should have some idea 
about tensors, which are mathematical objects used to describe the anisotropic properties of systems \cite{Feynman-Lectures-vol-2}. 
Taking into account that in Chapter 30 of Ref. \cite{Feynman-Lectures-vol-2} Feynman discusses the different properties of crystalline substances in distinct directions, in Chapter 31 \cite{Feynman-Lectures-vol-2} he presents a didactic approach to introduce the description of tensors, using the dependence of the induced polarization of a crystal on the direction of the applied electric field, as the basic example of an object with anisotropic property.
He also presents, in Chapter 31 \cite{Feynman-Lectures-vol-2}, 
the energy ellipsoid as a way of visualizing the polarization tensor. 

In the present paper, we propose 
an introduction to the notion of tensors, 
inspired by the aforementioned Feynman didactic approach,
found in Sections 31-1 to 31-3 of Ref. \cite{Feynman-Lectures-vol-2},
but considering a single ground state atom and a $\text{CO}_{2}$ molecule, instead of a crystal, as the examples that guide the initial discussions.
Moreover, we propose a preliminary visual representation of tensors, 
based on the ideas of the Lamé stress ellipsoid \cite{Fung-1965},
instead of the energy ellipsoid (as done in Ref. \cite{Feynman-Lectures-vol-2}).
These proposed changes aim to reduce the prerequisites of physical and mathematical concepts
required to follow the discussion, as well as to facilitate visual representations.
%
For example, to deal with crystals, in Ref. \cite{Feynman-Lectures-vol-2} 
the polarization $\textbf{P}$ is used, which is dipole moment per unit volume
\cite{Griffiths-Electrodynamics-1999,Feynman-Lectures-vol-2}.
Here, to deal with a single atom or molecule, just the dipole moment $\textbf{p}$ is necessary.
As another example, the perception of the isotropic polarizability of an
atom with a spherically symmetric electron cloud is more direct
than that of a cubic crystal.
Therefore, the consideration of a single atom (to illustrate an isotropic situation), 
or of a $\text{CO}_{2}$ molecule (to illustrate an anisotropic one),
can simplify the visualization of the polarization properties, if compared to crystals.
About the visualization of a tensor, to deal with an energy ellipsoid
requires, as discussed in Ref. \cite{Feynman-Lectures-vol-2}, some notion of differential and integral calculus, and also ideas on the energy per unit volume required to polarize a crystal.
On the other hand, only vector algebra is required to deal with the Lamé ellipsoid.
In this way, the modifications proposed here
are presented as a didactic proposal requiring less prerequisites 
if compared to Feynman's original approach,
and can be used directly as a learning tool for students, 
as well as for teachers interested in preparing their own materials.

The paper is organized as follows. 
We start reviewing some basic concepts: in Sec. \ref{sec-coordinat-transf}, 
we discuss coordinate transformation; in Sec. \ref{sec-scalar}, 
scalars; in Sec. \ref{sec-vector}, vectors.
In Sec. \ref{isotropic}, we discuss the polarizability of an isotropic particle,
taking a ground-state atom as example.
In Sec. \ref{polarizability-anisotropic-particle}, we discuss the polarizability of an anisotropic particle,
considering a $\text{CO}_{2}$ molecule as our basic model:
in Sec. \ref{polarizability-tensor-CO2}, we discuss the polarizability tensor;
in Sec. \ref{representing-tensor}, 
a diagonal matrix representation of this tensor;
in Sec. \ref{visulization}, 
a visual representation of this tensor, based on the
on the ideas of the Lamé stress ellipsoid;
in Sec. \ref{rotated-molecule-visual},
a visual representation of the polarizability tensor for a rotated molecule;
in Sec. \ref{rotated-molecule-matrix-representation},
a non-diagonal matrix representation of the tensor for a rotated molecule;
in Sec. \ref{rotated-system}, we return to the oriented molecule as discussed
in Secs. \ref{representing-tensor} and \ref{visulization}, and discuss the non-diagonal representation of the polarization tensor, now in a rotated coordinate system.
In Sec. \ref{sec:summary} we make a brief summary
of the main ideas discussed in this paper.
Finally, in Sec. \ref{sec-final}, we present our final comments.

\section{Basic ideas: coordinate transformation}
\label{sec-coordinat-transf}

\begin{figure}[h]
\centering  
\epsfig{file=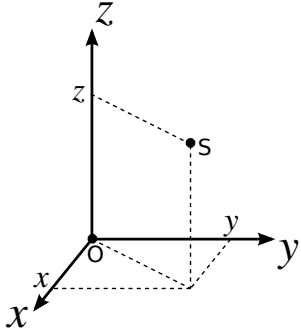,width=0.45 \linewidth}
		\caption{\footnotesize{
The points $O$ and $S$, and the Cartesian coordinate system $xyz$, whose origin coincides with $O$.
		}}
\label{fig:O-P-xyz}
\end{figure}

\begin{figure}[h]
\centering  
\subfigure[]{\label{fig:O-P-xyz-primo}\epsfig{file=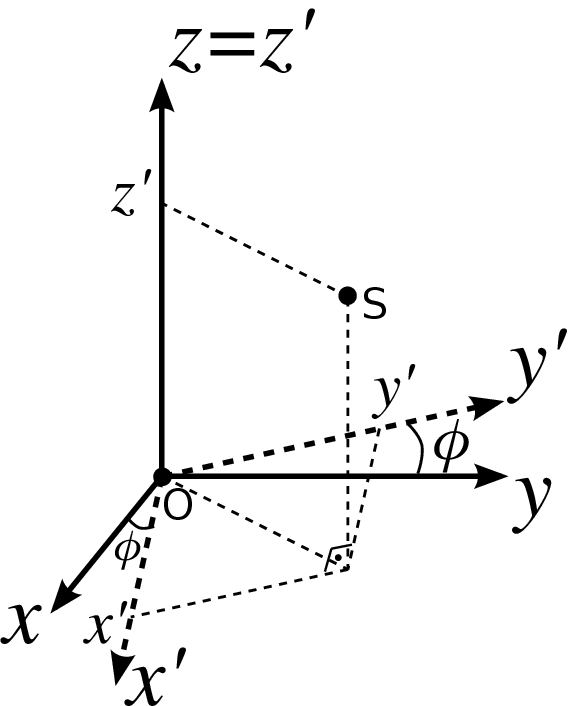,width=0.45 \linewidth}}
\hspace{5mm}
\subfigure[]{\label{fig:O-P-xyz-primo-planar}\epsfig{file=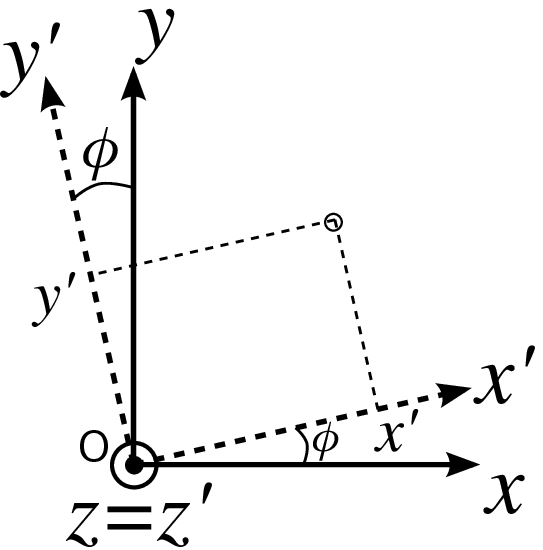,width=0.45 \linewidth}}
\caption{\footnotesize{
(a) Two points, $O$ and $S$, and two Cartesian coordinate systems: $xyz$, and $x^\prime y^\prime z^\prime$, being the latter rotated with respect to the former by an angle $\phi$, keeping the axis $z^\prime$ coinciding with $z$.
(b) Another visualization of the system $x^\prime y^\prime z^\prime$, rotated with respect to $xyz$, illustrated in Fig. (a). 
Here the axis $z^\prime=z$ appears perpendicular to the paper.
}}
\label{O-P-xyz-primo}
\end{figure}

Let us consider two points, $O$ and $S$, in space.
%
Considering a Cartesian coordinate system $xyz$, whose origin coincides, for convenience, with $O$, the points $O$ and $S$ are located by the coordinates $(0,0,0)$ and $(x,y,z)$, respectively [see Fig. \ref{fig:O-P-xyz}].
Through the present text, we are discussing how certain quantities described in a given coordinate system become described in another rotated one (specifically, we use the correlation between these descriptions to differentiate scalar, vectors and tensors).
Then, let us consider another Cartesian coordinate system $x^\prime y^\prime z^\prime$, rotated with respect to $xyz$ as illustrated in Fig. \ref{O-P-xyz-primo}, describing the same points 
$O$ and $S$ by the coordinates $(0,0,0)$ and $(x^\prime,y^\prime,z^\prime)$, respectively.
\begin{figure}[h]
\centering
\epsfig{file=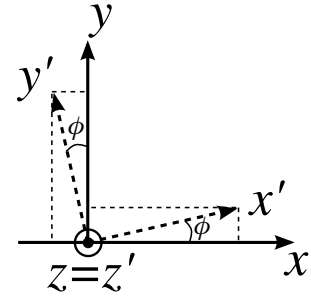,width=0.45 \linewidth}
\caption{\footnotesize{
Illustration of the coordinate systems $x^\prime y^\prime z^\prime$ and
$xyz$, which are related by Eq. \eqref{eq-x-primo-x-particular}.
}}
\label{fig:O-P-xyz-primo-relacao}
\end{figure}
Naturally, the point $S$ itself has not been changed,
but its description in  $xyz$ is different from that in $x^\prime y^\prime z^\prime$.
The relation between the coordinates $(x^\prime,y^\prime,z^\prime$) and
$(x,y,z)$ are given by (see Fig. \ref{fig:O-P-xyz-primo-relacao}):
\begin{eqnarray}
	x^\prime &=& \cos\phi\;x  + \sin\phi\;y,
	\\
	y^\prime &=& -\sin\phi\;x +  \cos\phi\;y,
	\\
	z^\prime &=& z.
	\label{eq:transf-0}
\end{eqnarray}
These relations can be written in matrix notation as
\begin{equation}
	\left[\begin{array}{c}
		x^\prime\\
		y^\prime\\
		z^\prime
	\end{array}\right]=\left[\begin{array}{ccc}
		\cos\phi& \sin\phi & 0\\
		-\sin\phi & \cos\phi & 0\\
		0 & 0 & 1
	\end{array}\right]\left[\begin{array}{c}
		{x}\\
		{y}\\
		{z}
	\end{array}\right]. \label{eq-x-primo-x-particular}
\end{equation}
Defining the square matrix in this equation as
\begin{equation}
\mathbf{R}^{(0)}=\left[\begin{array}{ccc}
\cos\phi& \sin\phi & 0\\
-\sin\phi & \cos\phi & 0\\
0 & 0 & 1
\end{array}\right], 
\label{eq-R0-def}
\end{equation}
one can note that 
\begin{equation}
	\mathbf{R}^{(0)}{\mathbf{R}}^{(0)T}=\mathbf{1},
	\label{eq-orto-0}
\end{equation}
where the superscript $T$ represents the transpose of the matrix, and
$\mathbf{1}$ is the $3\times 3$ identity matrix.
Eq. \eqref{eq-x-primo-x-particular} can also be written in a more compact manner,
using the index notation, as
\begin{equation}
	x^{\prime}_i=\sum_{r=1}^3{R}^{(0)}_{ir} x_r, \label{eq-x-primo-x-index-0}
\end{equation}
where $(x_1,x_2,x_3)=(x,y,z)$, $(x^{\prime}_1,x^{\prime}_2,x^{\prime}_3)=(x^{\prime},y^{\prime},z^{\prime})$, 
and ${R}^{(0)}_{ir}$ are the elements of the matrix $\mathbf{R}^{(0)}$.
In index notation, Eq. \eqref{eq-orto-0} can be rewritten as
\begin{equation}
	\sum_{k=1}^3 R^{(0)}_{ik}R^{(0)}_{jk}=\delta_{ij},
	\label{eq-delta-0}
\end{equation}
where $\delta_{ij}$ is the Kronecker delta symbol, defined 
by \cite{Butkov}
\begin{equation}
	\delta_{ij}=\begin{cases}
		0 & (\text{if \ensuremath{i\neq j}}),\\
		1 & (\text{if \ensuremath{i=j}}).
	\end{cases}
\end{equation}

\begin{figure}[h]
	\centering
	\epsfig{file=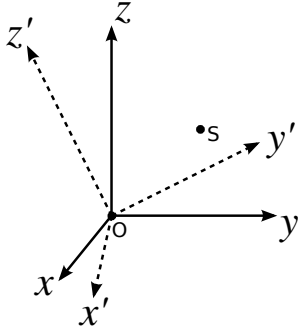,width=0.45 \linewidth}
	\caption{\footnotesize{
			General rotation of the system $x^\prime y^\prime z^\prime$  
			with respect to $xyz$.
	}}
	\label{rotacao-geral}
\end{figure}

For a general rotated Cartesian coordinate system $x^\prime y^\prime z^\prime$ (as illustrated in Fig. \ref{rotacao-geral}), 
the relation between the coordinates $(x^\prime,y^\prime,z^\prime$) and
$(x,y,z)$ takes the form \cite{Griffiths-Electrodynamics-1999}
\begin{equation}
	\left[\begin{array}{c}
		x^\prime\\
		y^\prime\\
		z^\prime
	\end{array}\right]=\left[\begin{array}{ccc}
		R_{11}& R_{12} & R_{13}\\
		R_{21} & R_{22} & R_{23}\\
		R_{31} & R_{32} & R_{33}
	\end{array}\right]\left[\begin{array}{c}
		{x}\\
		{y}\\
		{z}
	\end{array}\right], \label{eq-x-primo-x}
\end{equation}
where $R_{ij}$ are the elements of a matrix $\textbf{R}$,
%
which can be written, in terms of the Euler angles,
as shown in Appendix \ref{ap:R}. 
The explicit form of this matrix is just for informational purposes, since it is not necessary to follow the reasoning through this article.
However, an important feature of $\textbf{R}$, relevant to the present discussion, is that it is orthogonal, which means that
\begin{equation}
	\mathbf{R}\mathbf{R}^T=\mathbf{1}.
	\label{eq-orto}
\end{equation}
Note that $\textbf{R}$ is a generalization of the matrix $\textbf{R}^{(0)}$,
and Eq. \eqref{eq-orto} is a generalization of Eq. \eqref{eq-orto-0}.
Eqs. \eqref{eq-x-primo-x} and \eqref{eq-orto} can be written in index notation, respectively,
by
\begin{equation}
	x^{\prime}_i = \sum_{j=1}^3 R_{ij} x_j, 
	\label{eq-x-primo-x-index}
\end{equation}
and
\begin{equation}
	\sum_{k=1}^3 R_{ik}R_{jk}=\delta_{ij}.
	\label{eq-delta}
\end{equation}

\section{Basic ideas: scalars}
\label{sec-scalar}

From the point of view of the system $xyz$, the distance from $O$ to $S$ is visually
represented by a line segment,
and numerically represented
by a number $d$, which is given, in terms of the coordinates and in index notation, by
\begin{equation}
	d^2=\sum_{i=1}^3 x_i x_i. \label{d-index}
\end{equation}
Considering the system $x^\prime y^\prime z^\prime$, the distance 
from $O$ to $S$, represented by the number $d^\prime$ 
(everything in this system we indicate by the superscript "$\prime$"), is given by
\begin{equation}
	d^{\prime 2}=\sum_{i=1}^3x^{\prime}_i x^{\prime}_i. \label{d-index-primo}
\end{equation}
Using Eqs. \eqref{eq-x-primo-x-index-0}, \eqref{eq-delta-0}, and \eqref{d-index} in Eq. \eqref{d-index-primo}, we obtain 
\begin{equation}
	d^{\prime 2}=d^{2}. \label{d-d-primo}
\end{equation}
Then, the distance from $O$ to $S$ is represented by the same number in both 
coordinate systems $xyz$ and $x^\prime y^\prime z^\prime$, as naturally expected.
The distance is an example of a scalar quantity, 
in the sense that it is represented by a number invariant under a coordinate transformation, as that given in Eq. \eqref{eq-x-primo-x-particular}.
In other words, a scalar is a quantity characterized by just one number,
which is independent of the coordinate system we are using to describe this quantity \cite{Fleisch-StudentsGuide-2011,Arfken-MathematicalMethods-2005};
or, according to Feynman \cite{Feynman-Lectures-vol-2}, 
\begin{quote}
	\textit{``... a number independent of the choice of axes''.} 
\end{quote}

For a general rotated Cartesian coordinate system $x^\prime y^\prime z^\prime$ (as illustrated in Fig. \ref{rotacao-geral}), using Eqs. \eqref{eq-x-primo-x-index} and \eqref{eq-delta} in 
Eq. \eqref{d-index-primo},
we obtain again that $d=d^\prime$, so that the
distance from $O$ to $S$ is represented in the same manner in
the system $xyz$, or in any rotated system $x^\prime y^\prime z^\prime$. 
In other words, the distance, or the value that represents it, does not
change under the operation of a rotation of the coordinate system, or is a symmetry
or symmetrical under this operation.
Symmetry is a key concept in understanding the laws of physics \cite{Feynman-Lectures-vol-1}.
According to Feynman in Sec. 11-1 in Ref. \cite{Feynman-Lectures-vol-1},
\begin{quote}
	\textit{``Professor Hermann Weyl has given this definition of symmetry: a thing is symmetrical if one can subject it to a certain operation and it appears exactly the same after the operation.''}
\end{quote}

In this section, we considered the distance between two points as our base example of a scalar, but  other examples of scalar quantities include temperature, energy, mass, charge, among others.

\section{Basic ideas: vectors}
\label{sec-vector}

\begin{figure}[h]
\centering  
\subfigure[]{\label{fig:duas-cargas}\epsfig{file=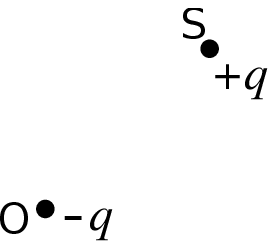,width=0.25 \linewidth}}
\hspace{2cm}
\subfigure[]{\label{fig:duas-cargas-p}\epsfig{file=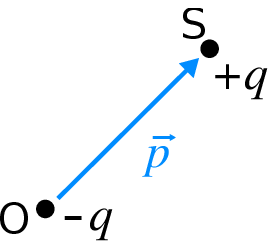,width=0.25 \linewidth}}
\caption{\footnotesize{
(a) A dipole formed by a point charge $-q$ (with $q>0$) at the point $O$, and $+q$ at $S$.
(b) The dipole moment vector $\textbf{p}$, which characterizes the dipole formed by these two point charges.
}}
\label{duas-cargas}
\end{figure}

\begin{figure}[h]
\centering
		\epsfig{file=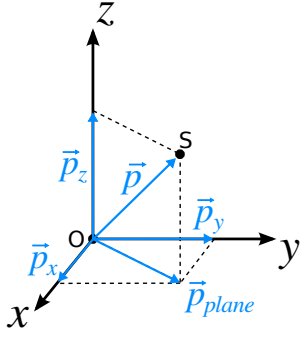,width=0.45 \linewidth}
		\caption{\footnotesize{
				The Cartesian coordinate system $xyz$, whose origin coincides
				with $O$, and the dipole moment vector $\textbf{p}$, 
				which characterizes the dipole formed by the charges $-q$ (with $q>0$) and $+q$.
				}}
		\label{duas-cargas-p-xyz}
\end{figure}

\begin{figure}[h]
\centering
		\epsfig{file=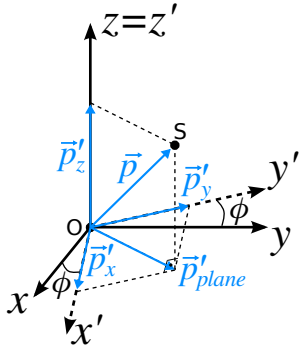,width=0.45 \linewidth}
		\caption{\footnotesize{
				The dipole moment vector $\textbf{p}$, and two Cartesian coordinate systems: $xyz$,
				and $x^\prime y^\prime z^\prime$, being the latter rotated 
				with respect to the former by an angle $\phi$, keeping the axis $z^\prime$
				coinciding with $z$. 				
			}}
		\label{duas-cargas-p-xyz-primo}
\end{figure}

\begin{figure}[h]
\centering
		\epsfig{file=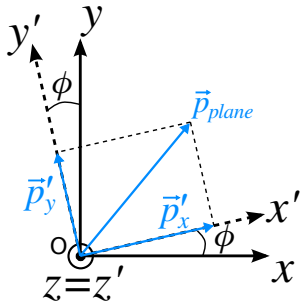,width=0.45 \linewidth}
		\caption{\footnotesize{
				Another visualization of the system $x^\prime y^\prime z^\prime$ rotated 
				with respect to $xyz$, illustrated in Fig. \ref{duas-cargas-p-xyz-primo}. 
				Here the axis $z^\prime=z$ appears perpendicular to the paper.				
		}}
		\label{duas-cargas-p-xyz-primo-planar}
\end{figure}

Let us consider again two points, $O$ and $S$, in space [Fig. \ref{fig:O-P-xyz}].
Let us imagine a point charge $q>0$ at $S$, and another one $-q$ at $O$ [see Fig. \ref{fig:duas-cargas}], so that, they form an electric dipole. 
A visual representation of this dipole can be done by an arrow $\textbf{p}$ (or also $\vec{p}$), called dipole moment vector, pointing from the negative to the positive charge (this is a convention) \cite{Griffiths-Electrodynamics-1999}, whose length is directly proportional to the product $qd$ ($d$ is the distance between these charges), as illustrated in Fig. \ref{fig:duas-cargas-p}.  
When we consider a Cartesian coordinate system $xyz$,
$\textbf{p}$ is described by the components $(p_x, p_y, p_z)$ (see Fig. \ref{duas-cargas-p-xyz}). 
Another Cartesian coordinate system $x^\prime y^\prime z^\prime$, rotated with respect to $xyz$, as illustrated in Fig. \ref{duas-cargas-p-xyz-primo}, describes $\textbf{p}$ by $(p_{x}^\prime,p_{y}^\prime,p_{z}^\prime)$.
Note that the vector $\textbf{p}$ itself has not been changed,
but its description in  $xyz$ is different from that in $x^\prime y^\prime z^\prime$.
The relation between the components $(p_{x}^\prime,p_{y}^\prime,p_{z}^\prime)$ and
$(p_x,p_y,p_z)$ are given by (see Fig. \ref{duas-cargas-p-xyz-primo-planar}):
\begin{eqnarray}
	p_{x}^\prime &=& \cos\phi\;  p_{x} + \sin\phi\; p_{y},
	\\
	p_{y}^\prime &=& -\sin\phi\;  p_{x} + \cos\phi\; p_{y},
	\\
	p_{z}^\prime &=&  p_{z}.
\end{eqnarray}
These relations can be written in matrix notation as
\begin{equation}
	\left[\begin{array}{c}
		p_{x}^\prime\\
		p_{y}^\prime\\
		p_{z}^\prime
	\end{array}\right]=\left[\begin{array}{ccc}
		\cos\phi& \sin\phi & 0\\
		-\sin\phi & \cos\phi & 0\\
		0 & 0 & 1
	\end{array}\right]\left[\begin{array}{c}
		p_{x}\\
		p_{y}\\
		p_{z}
	\end{array}\right], 
\label{eq-px-primo-px-particular}
\end{equation}
and in index notation as
\begin{equation}
	p^{\prime}_i=\sum_{r=1}^3{R}^{(0)}_{ir} p_r, \label{eq-p-primo-p-particular-indices}
\end{equation}
where $(p_1,p_2,p_3)=(p_x,p_y,p_z)$ and $(p^{\prime}_1,p^{\prime}_2,p^{\prime}_3)=(p_x^{\prime},p_y^{\prime},p_z^{\prime})$.
%
Note that the square matrix in this equation is the same found in Eq. \eqref{eq-x-primo-x-particular}.
This means that the components of $\textbf{p}$ transform in the same way that the coordinates of a point in space.
We can say that the components of a general vector $\textbf{v}$, under the rotation of the coordinate system shown in Fig. \ref{fig:O-P-xyz-primo-relacao} and described by Eq. \eqref{eq-x-primo-x-particular}, transform like the coordinates $(x,y,z)$, so that  
\begin{equation}
	v_{i}^{\prime} = \sum_{r=1}^3 R^{(0)}_{ir} v_{r}. \label{eq-v-primo-v-index-R0}
\end{equation}
In this context, a three-dimensional vector is a set of three quantities which transform, under a rotation of the coordinate system, in the same manner that the three-dimensional coordinates of a point in space  \cite{Landau-Lifshitz-vol-2}. 

\begin{figure}[h]
\centering
		\epsfig{file=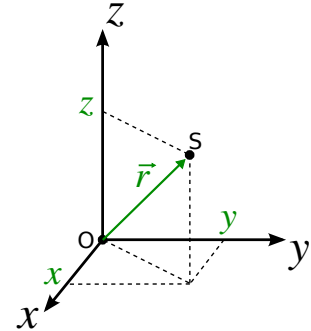,width=0.45 \linewidth}
		\caption{\footnotesize{
		The position vector $\textbf{r}$, whose componets are the coordinates $(x,y,z)$.
		}}
		\label{r-vetor}
\end{figure}

A vector is a mathematical object characterized by a magnitude and a direction associated with it \cite{Fleisch-StudentsGuide-2011, Arfken-MathematicalMethods-2005}.
The coordinates of a point in space, can be considered, themselves, as components of a vector $\textbf{r}$ named as position vector (see Fig. \ref{r-vetor}) \cite{Landau-Lifshitz-vol-2}.

For a general rotated Cartesian coordinate system $x^\prime y^\prime z^\prime$ 
(as illustrated in Fig. \ref{rotacao-geral}), 
the relation between the components $(v_{x}^\prime,v_{y}^\prime,v_{z}^\prime$) and
$(v_x,v_y,v_z)$, of a general vector $\textbf{v}$, takes the form
\begin{equation}
	v_{i}^{\prime} = \sum_{r=1}^3 R_{ir} v_{r}. \label{eq-v-primo-v-index}
\end{equation}
This means that the components of $\textbf{v}$ transform in the same way that the coordinates in Eq. \eqref{eq-x-primo-x-index}.
\begin{figure}[h]
\centering
\epsfig{file=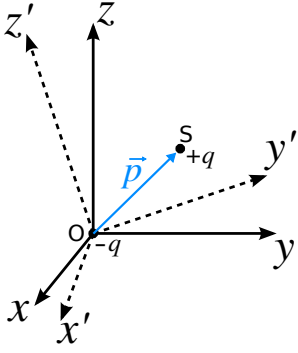,width=0.45 \linewidth}
\caption{\footnotesize{
General rotation of the system $x^\prime y^\prime z^\prime$,  
with respect to $xyz$. It is also shown the dipole moment vector $\textbf{p}$.
}}
\label{duas-cargas-p-xyz-primo-geral}
\end{figure}

From the point of view of the system $xyz$, the magnitude of the vector $\textbf{v}$, here called $v$, 
is such that
\begin{equation}
	v^2=v_x^2+v_y^2+v_z^2=\sum_{i=1}^3 v_i v_i. \label{v-index-0}
\end{equation}
From the point of view of the system $x^\prime y^\prime z^\prime$, the magnitude
of $\textbf{v}$ is such that
\begin{equation}
	v^{\prime 2}=\sum_{i=1}^3 v^{\prime}_i v^{\prime}_i. \label{v-index-1}
\end{equation}
Using \eqref{eq-v-primo-v-index} and \eqref{eq-delta} in \eqref{v-index-1},
we obtain that
\begin{equation}
	v^{\prime}=v.
	\label{eq-v-scalar}
\end{equation}
In other words, the magnitude of a vector $\textbf{v}$ is a scalar.

According to Feynman (Chap. 11 in Ref.  \cite{Feynman-Lectures-vol-1}),
a vector is
\begin{quote}
	\textit{``A ``directed quantity'' (which is really 3 quantities; components $a_x$, $a_y$, $a_z$
	on three axes)... represented by a single symbol $\vec{a}$''}.
\end{quote}
(Here, we are using, as a choice, the notation $\textbf{a}$ instead of $\vec{a}$.)
There are some basic operations which involve vectors. 
First, two vectors $\textbf{u}$ and $\textbf{v}$ can be added
\cite{Feynman-Lectures-vol-1,Griffiths-Electrodynamics-1999},
\begin{equation}
	\textbf{u}+\textbf{v}=\textbf{v}+\textbf{u}=\textbf{w},
\end{equation}
so that, the components of $\textbf{w}$ are
%
%
\begin{equation}
	{w_i}={u_i}+{v_i}.
\end{equation}
Another basic operation with vectors is the multiplication by a scalar $\mu$,
\begin{equation}
	\mu\textbf{v}=\textbf{w},
\end{equation}
which means
\begin{equation}
	{w_i}=\mu{v_i}.
\end{equation}
From two vectors, we can also build a scalar by means of an operation called scalar product, defined by
\begin{equation}
	\textbf{v}\cdot\textbf{w}=\sum_{i=1}^{3}v_iw_i,
	\label{eq-p-scalar}
\end{equation}
which, from a geometrical point of view is given by
\begin{equation}
	\textbf{v}\cdot\textbf{w}= v w \cos(\theta),
	\label{eq-scalar-geom}
\end{equation}
where $\theta$ is the angle between the vectors.
%
Note that, using \eqref{eq-v-primo-v-index} and \eqref{eq-delta},
we obtain that $\sum_{i=1}^{3}v_i^{\prime}w_i^{\prime}=\sum_{i=1}^{3}v_iw_i$, so that
the result of the product $\textbf{v}\cdot\textbf{w}$ 
is an invariant under rotations of the coordinate system, which characterizes it as a scalar \cite{Landau-Lifshitz-vol-2}.

The unit vectors  $\hat{\textbf{x}}$, $\hat{\textbf{y}}$
and $\hat{\textbf{z}}$, pointing to the $x$, $y$, and $z$ directions, respectively, 
form a basis, so that any vector $\textbf{y}$ can be written as
\begin{equation}
	\textbf{v}=v_1\hat{\textbf{x}}+v_2 \hat{\textbf{y}}
	+v_3\hat{\textbf{z}},
	\label{eq:v-expanded}
\end{equation}
where $v_1=\textbf{v}\cdot\hat{\textbf{x}}$,
$v_2=\textbf{v}\cdot\hat{\textbf{y}}$,
and $v_3=\textbf{v}\cdot\hat{\textbf{z}}$.
Using the notation $\hat{\mathbf{e}}_1=\hat{\textbf{x}}$, $\hat{\mathbf{e}}_2=\hat{\textbf{y}}$
and $\hat{\mathbf{e}}_3=\hat{\textbf{z}}$, one can write
\begin{equation}
	\textbf{v}=\sum_{i=1}^3 v_i\hat{\mathbf{e}}_i.
	\label{eq:v-expanded-indice}
\end{equation}

In this section, we discussed vectors,
focusing, as base examples, on the electric dipole moment and position vectors.
Other examples of vector quantities include force, velocity, acceleration, among others
\cite{Fleisch-StudentsGuide-2011}.

\section{Polarizability of an isotropic particle}
\label{isotropic}

\begin{figure}[h]
\centering  
\subfigure[]{\label{fig:atomo1}\epsfig{file=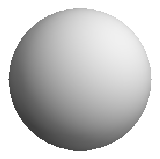, width=0.3 \linewidth}}
\hspace{2mm}
\subfigure[]{\label{fig:atomo2}\epsfig{file=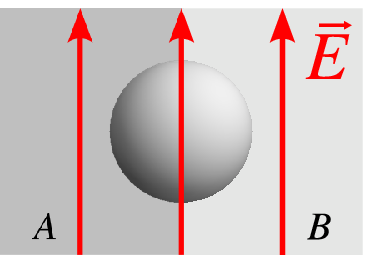, width=0.3 \linewidth}}
\hspace{2mm}
\subfigure[]{\label{fig:atomo3}\epsfig{file=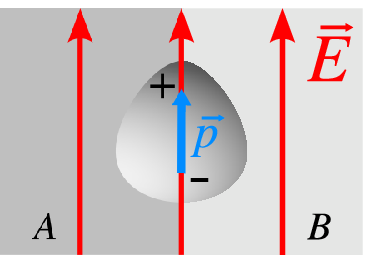, width=0.3 \linewidth}}
\caption{\footnotesize{ 
(a) Representation of the electron distribution around the nucleus of an atom in the ground state.
Note that the electron distribution has spherical symmetry.
(b) Representation of the atom in the presence of an external uniform electric field $\textbf{E}$.
Note that the structure of the atom in region $A$ (dark), is identical to that in region $B$ (lighter).
(c) Representation of the electron distribution around the nucleus of an atom in the ground state, in the presence of $\textbf{E}$.
Note that the symmetry between the regions $A$ and $B$ remains, so that the induced dipole moment $\textbf{p}$ has the same direction than $\textbf{E}$.
}}
\label{atomo}
\end{figure}

\begin{figure}
\centering  
\subfigure[]{\label{fig:atomo-lab-rotacao}\epsfig{file=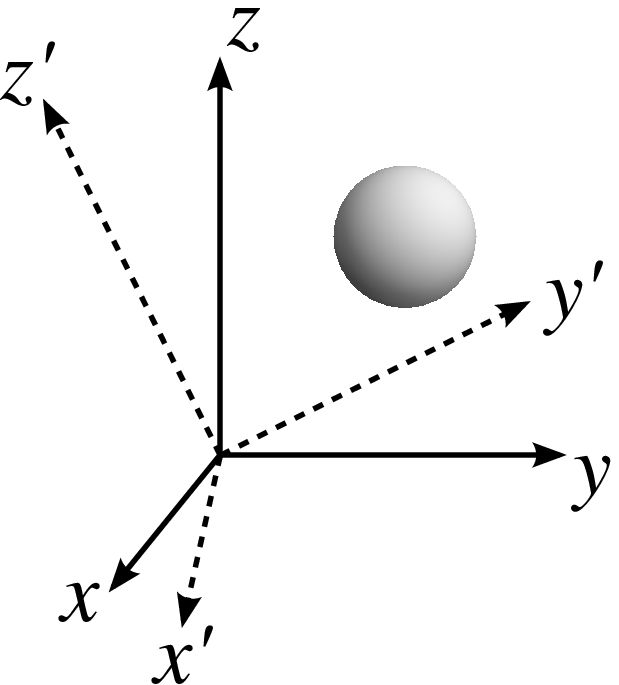, width=0.45 \linewidth}}
\hspace{3mm}
\subfigure[]{\label{fig:atomo-rotacao}\epsfig{file=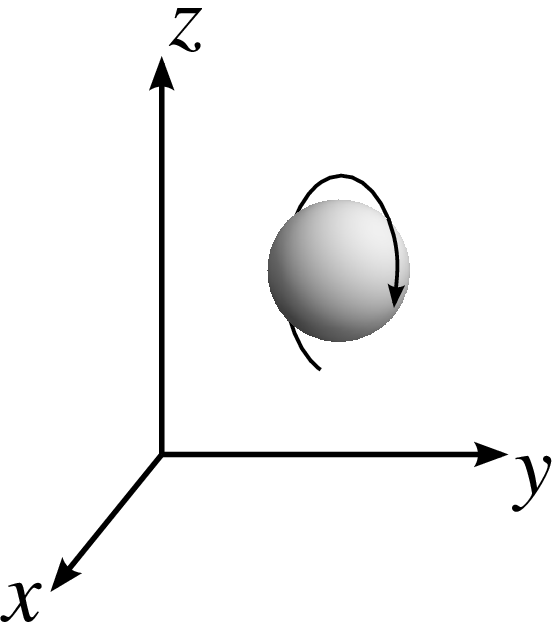, width=0.45 \linewidth}}
\caption{\footnotesize{ 
(a) General rotation of the system $x^\prime y^\prime z^\prime$, with respect to $xyz$.
It is also shown an atom in the ground state, so that, its electronic distribution has spherical symmetry.
(b) The atom rotated with respect to a coordinate system $xyz$. 
}}
\end{figure}

According to Feynman \cite{Feynman-Lectures-vol-2},
\begin{quote}
	``\textit{... in physics we usually start out by talking about the special case in which the polarizability is the same in all 
	directions, to make life easier.''}
\end{quote}
Following this comment, let us start considering the case where the polarizability is the same in all directions, in other words, the case where this property is called isotropic.
In this way, we consider a neutral atom in its ground state.
For any atom, no matter how many electrons it contains, in its ground state the electron distribution around its nucleus has spherical symmetry \cite{Purcell-Electrodynamics-2011}, as illustrated in Fig. \ref{fig:atomo1}.
Let us also consider this neutral atom in the presence of an external uniform electric field $\textbf{E}$, as illustrated in Fig. \ref{fig:atomo2}.
When $\textbf{E}$ is applied, the positive nucleus of the atom is pushed in the direction of the field, whereas, its electrons are pulled in the opposite direction, so that the atom becomes polarized [see Fig. \ref{fig:atomo3}] \cite{Griffiths-Electrodynamics-1999}.
[Note that the electron cloud becomes deformed (see, for instance, Ref. \cite{Purcell-Electrodynamics-2011}).]
Thus the electric field $\textbf{E}$ induces in the atom a dipole moment $\textbf{p}$.
Note in Fig. \ref{fig:atomo3} that the structure of the atom (with the deformed electron cloud) in region $A$, is identical to that in region $B$, so that, no matter what is the direction of $\textbf{E}$, by symmetry arguments it is not expected that $\textbf{p}$ points to other direction than parallel to $\textbf{E}$.
In fact, the dipole moment $\textbf{p}$, in this case, is given by
\begin{equation}
	\text{\textbf{p}}= \alpha {\textbf{E}},
	\label{p-beta-E}
\end{equation}
where $\alpha$ is the atomic polarizability \cite{Griffiths-Electrodynamics-1999},
which establishes the connection between the induced dipole moment $\textbf{p}$
and an incident field $\textbf{E}$. 
Note that $\alpha$ is a scalar, which means that Eq. \eqref{p-beta-E} remains valid if a coordinate system is rotated with respect to the atom [see Fig. \ref{fig:atomo-lab-rotacao}], or vice-versa [see Fig. \ref{fig:atomo-rotacao}].

\section{Polarizability of an anisotropic particle}
\label{polarizability-anisotropic-particle}

\subsection{Polarizability tensor for a $\text{CO}_{2}$ molecule}
\label{polarizability-tensor-CO2}

\begin{figure}
\centering 
\subfigure[]{\label{fig:molecula}\epsfig{file=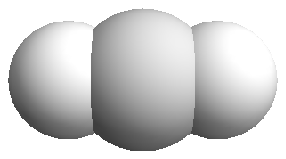, width=0.4 \linewidth}}
\hspace{10mm}
\subfigure[]{\label{fig:molecula-simetria-paral}\epsfig{file=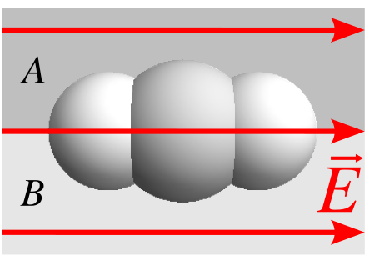, width=0.4 \linewidth}}
\subfigure[]{\label{fig:molecula-simetria-perp}\epsfig{file=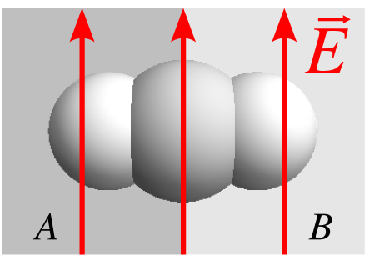, width=0.4 \linewidth}}
\hspace{10mm}
\subfigure[]{\label{fig:molecula-eixos-principais}\epsfig{file=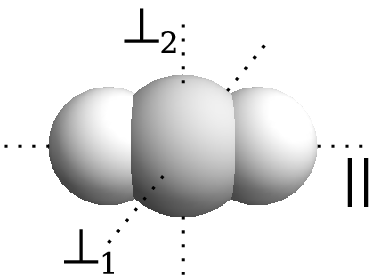, width=0.4 \linewidth}}
\caption{\footnotesize{ 
(a) Representation of the electron distribution of a $\text{CO}_{2}$ molecule in the ground state. 
Note that, different from a single atom, the electron distribution of this molecule does not have a spherical symmetry. 
In (b) and (c), an uniform electric field $\textbf{E}$ (represented by the arrow) is applied parallel (b), and perpendicular (c) to the molecule axis.
Note that the structure of the $\text{CO}_{2}$ molecule in region A (dark), is identical to that in region B (lighter), so that, by the symmetry between the regions A and B, it is not expected that the dipole moment vector points to other direction than the direction of $\textbf{E}$, which is expressed in Eq. \eqref{eq-p-prop-E-axis-CO2} [for (b)], and \eqref{eq-p-prop-E-perp-CO2} [for (c)].
(d) A set of two directions perpendicular (indicated by $\bot_{1}$ and $\bot_{2}$), and the one parallel (indicated by $\parallel$) to the molecule axis.
}}
\end{figure}

When we have different polarizability in different directions, this property is called anisotropic.
In order to discuss this case, we consider a $\text{CO}_{2}$ molecule, as illustrated in Fig. \ref{fig:molecula}.
Note that, different from the atom considered before, this molecule does not have a spherical symmetry. 
Despite this, it has a symmetry axis, named the molecule axis \cite{Griffiths-Electrodynamics-1999}, which is the one that crosses the nuclei of the atoms that form the molecule.

Let us consider the $\text{CO}_{2}$ molecule in the presence of an external uniform electric field $\textbf{E}$.
In Fig. \ref{fig:molecula-simetria-paral}, one can see the case where $\textbf{E}$ is applied parallel to the molecule axis. 
In this figure, one can note that the structure of the $\text{CO}_{2}$ molecule in region $A$, is identical to that in region $B$, so that, by symmetry arguments, it is not expected that the dipole moment vector points to other direction than parallel to the molecule axis and in the same direction of $\textbf{E}$.
In fact, the dipole moment $\textbf{p}$ in this case, is given by
\begin{equation}
	\textbf{p} = \alpha_{\parallel} \textbf{E},
	\label{eq-p-prop-E-axis-CO2}
\end{equation}
where $\alpha_{\parallel}>0$
is the polarizability of the molecule in the direction of its axis (the subscript $\parallel$ refers to this direction).
When the electric field $\textbf{E}$ is applied in a direction perpendicular to the molecule axis, one can see in Fig. \ref{fig:molecula-simetria-perp} that, the structure of the $\text{CO}_{2}$ molecule in region $A$, is identical to that in region $B$, so that, by symmetry arguments, it is not expected that the dipole moment vector points to other direction than perpendicular to the molecule axis and in the same direction of $\textbf{E}$.
In fact, the dipole moment in this case, is given by
\begin{equation}
	\textbf{p} = \alpha_{\bot} \textbf{E},
	\label{eq-p-prop-E-perp-CO2}
\end{equation}
where $\alpha_{\bot}$
is the polarizability of the molecule in the directions perpendicular to its axis (the subscript $\perp$ refers to these directions).
The set of two directions perpendicular to the molecule axis, together with that parallel to this axis, as illustrated in Fig. \ref{fig:molecula-eixos-principais}, are known as the principal axes of the $\text{CO}_{2}$ molecule.
Moreover, when the electric field $\textbf{E}$ is applied in a direction not coinciding with one of these principal axes, we have that the polarization $\textbf{p}$ is no longer in the same direction as the electric field $\textbf{E}$ (this is discussed next).
Thus, the connection between the induced dipole moment $\textbf{p}$ 
and the applied electric field $\textbf{E}$ is more complex than that of an isotropic atom.
This connection between $\textbf{p}$ and  $\textbf{E}$ in this case is given by the polarizability tensor $\overleftrightarrow{\alpha}$ of the $\text{CO}_{2}$ molecule.
A physical quantity that is characterized by magnitudes, which are associated to multiple directions, is called a tensor \cite{Fleisch-StudentsGuide-2011}.
In the next section, we discuss a representation of $\overleftrightarrow{\alpha}$.

\subsection{Matrix representation of the polarizability tensor for a $\text{CO}_{2}$ molecule}
\label{representing-tensor}

\begin{figure}[h]
\centering 
\subfigure[]{\label{molecula-eixo-y}\epsfig{file=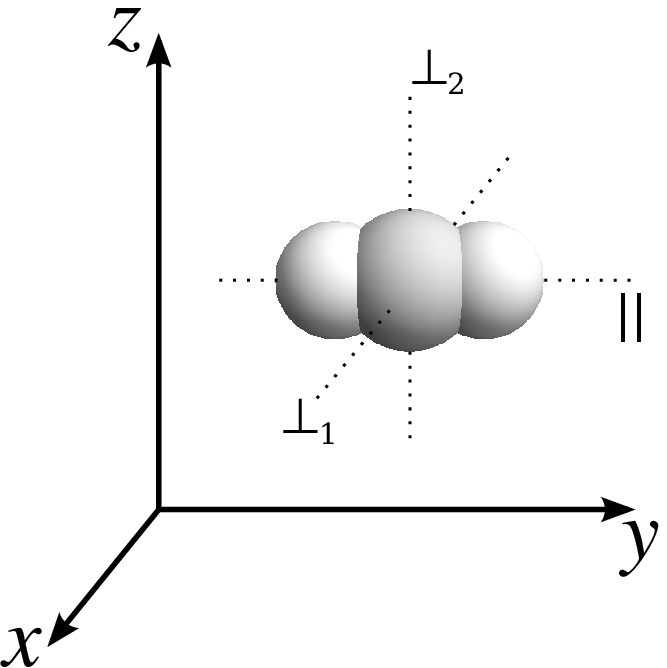, width=0.45 \linewidth}}
\hspace{5mm}
\subfigure[]{\label{fig:molecula-eixo-y-plano}\epsfig{file=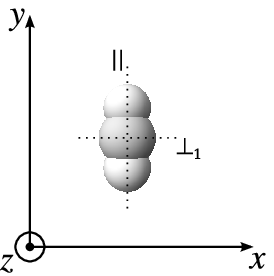, width=0.45 \linewidth}}
\caption{\footnotesize{ 
(a)	A $\text{CO}_{2}$ molecule, oriented in space in a such way that its principal axes 
(dotted lines) are parallel to those of the laboratory $xyz$-system.
(b) Another point of view, in which the $z$-axis appears perpendicular to the paper.
}}
\label{fig:molecula-eixo-y}
\end{figure}

\begin{figure}[h]
\centering  
\subfigure[]{\label{fig:molecula-campo-x}\epsfig{file=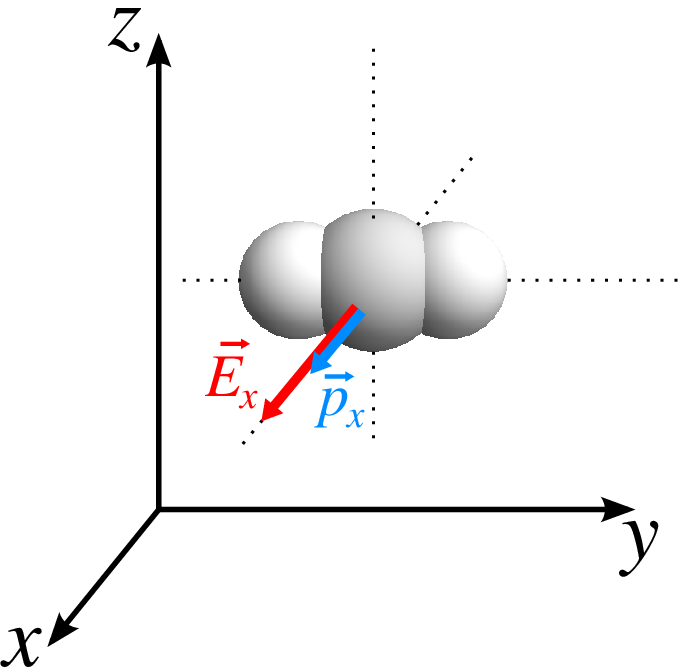, width=0.45 \linewidth}}
\hspace{2mm}
\subfigure[]{\label{fig:molecula-campo-y}\epsfig{file=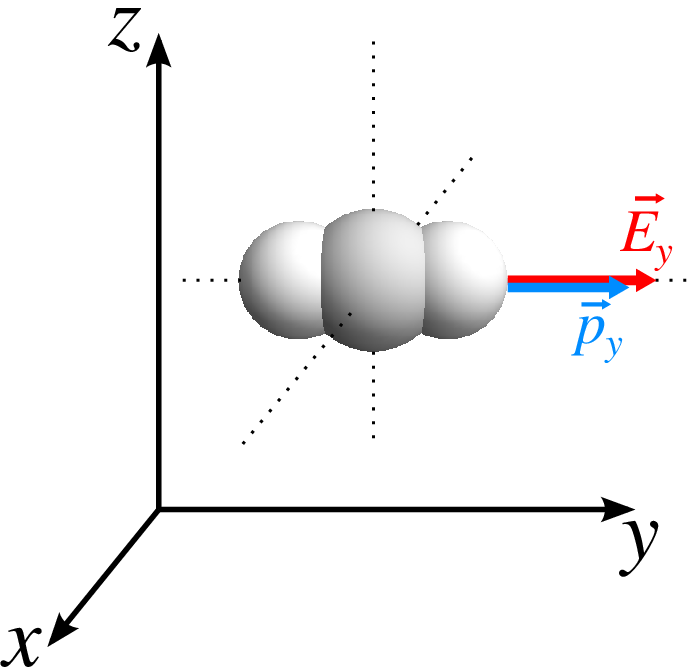, width=0.45 \linewidth}}
\subfigure[]{\label{fig:molecula-campo-z}\epsfig{file=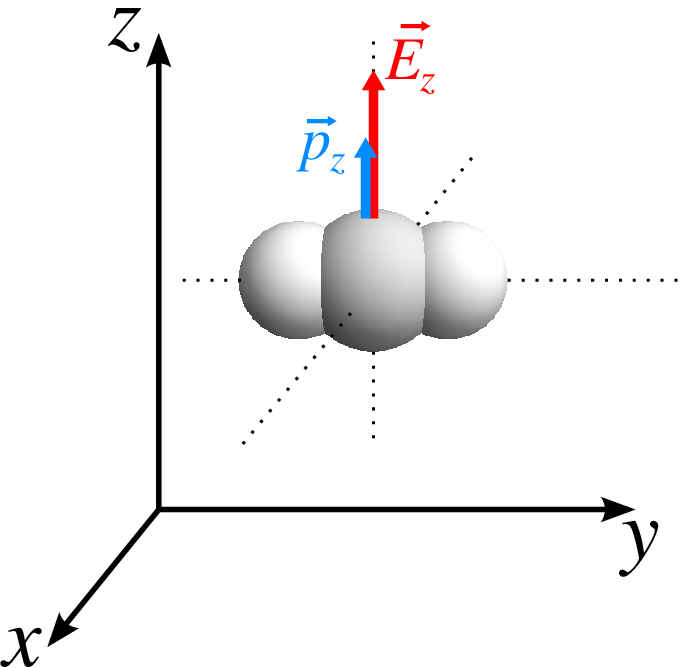, width=0.45 \linewidth}}
\caption{\footnotesize{ 
A $\text{CO}_{2}$ molecule, oriented in space in a such way that its principal axes 
(dotted lines) are parallel to those  of the laboratory $xyz$-system.
(a) An electric field $\textbf{E}_{x}$, in the $x$-direction, induces a dipole moment $\textbf{p}_{x}$, 
only in the $x$-direction.
(b) An electric field $\textbf{E}_{y}$, in the $y$-direction, induces a dipole moment $\textbf{p}_{y}$, 
only in $y$-direction.
(c) An electric field $\textbf{E}_{z}$, in the $z$-direction, induces a dipole moment $\textbf{p}_{z}$, 
only in $z$-direction.
}}
\end{figure}

\begin{figure}[h]
	\centering  
	\subfigure[]{\label{fig:molecula-campo-45}\epsfig{file=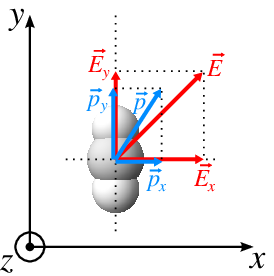, width=0.45 \linewidth}}
	\hspace{2mm}
	\subfigure[]{\label{fig:molecula-campo-60}\epsfig{file=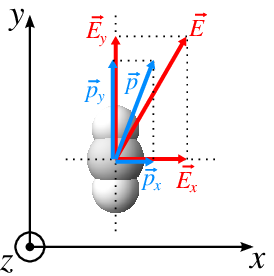, width=0.45 \linewidth}}
	\caption{
		\footnotesize{ 
			A $\text{CO}_{2}$ molecule, oriented in space in a such way that its principal axes 
			(dotted lines) are parallel to those  of the laboratory $xyz$-system.
			Note that the $z$-axis appears perpendicular to the paper.
			(a) A field $\textbf{E}$, applied at $45\degree$.
			(b) A field $\textbf{E}$, applied at $60\degree$.
			Note that, in (a) and (b), the induced dipole moment $\textbf{p}$ is not in the same direction of $\textbf{E}$.
		}}
\end{figure}

Let us start considering a laboratory Cartesian system $xyz$, and put the $\text{CO}_{2}$ molecule oriented in space in a such way that its principal axes are parallel to the $xyz$ axes, as shown in Fig. \ref{fig:molecula-eixo-y} (this is just a convenient choice to start our discussion).
When we apply an electric field $\textbf{E}_{x}$ in the $x$-direction, this produces, according to Eq. \eqref{eq-p-prop-E-perp-CO2}, an induced dipole moment $\textbf{p}_{x}$ only in the $x$-direction [see Fig. \ref{fig:molecula-campo-x}], given by
\begin{equation}
	\textbf{p}_{x}=\alpha_{\bot} \textbf{E}_{x}.
	\label{P1-CO2}
\end{equation}
When we apply an electric field $\textbf{E}_y$ in the $y$-direction, this produces, according to Eq. \eqref{eq-p-prop-E-axis-CO2}, an induced dipole moment $\textbf{p}_{y}$ only in the $y$-direction [see Fig. \ref{fig:molecula-campo-y}], given by
\begin{equation}
\textbf{p}_{y}=\alpha_{\parallel} \textbf{E}_{y}.
\label{P2-CO2}
\end{equation}
Lastly, when an electric field $\textbf{E}_z$ is applied in the $z$-direction, this produces an induced dipole moment $\textbf{p}_{z}$ only in the $z$-direction [see Fig. \ref{fig:molecula-campo-z}], given by
\begin{equation}
	\textbf{p}_{z}=\alpha_{\bot} \textbf{E}_{z}.
	\label{P3-CO2}
\end{equation}
To analyze a superposition of the fields $\textbf{E}_{x}$ and $\textbf{E}_{y}$,
let us consider Feynman's forwarding:
\begin{quote}
	\textit{``Suppose, in a particular crystal, we find that an electric field $\textbf{E}_1$ in the $x$-direction produces the polarization
		$\textbf{P}_1$ in the $x$-direction.
		Then we find that an electric field $\textbf{E}_2$ in the $y$-direction, with the same strength as $\textbf{E}_1$, produces a
		different polarization $\textbf{P}_2$ in the $y$-direction. What would happen if we put an electric field at $45\degree$?''}
\end{quote} 
This Feynman sentence refers to a crystal and a vector polarization $\textbf{P}_1$ (dipole moment per unit volume of
the crystal).
Here, we use his sentence, but adapting it to our case, by considering a $\text{CO}_{2}$ molecule, instead of a crystal, and replacing $\textbf{P}_1 \to\textbf{p}_1$.
In this way, the induced dipole moments $\textbf{p}_x$ and $\textbf{p}_y$, produced by the electric fields $\textbf{E}_x=E_x \hat{\textbf{x}}$ and $\textbf{E}_y=E_y \hat{\textbf{y}}$, respectively, are given by Eqs. \eqref{eq-p-prop-E-axis-CO2} and \eqref{eq-p-prop-E-perp-CO2}.
Then, if we apply an electric field $\textbf{E}=\textbf{E}_x+\textbf{E}_y$, the induced dipole moment $\textbf{p}$ will be the vector sum of $\textbf{p}_{x}$ and $\textbf{p}_{y}$, which is given by
\begin{equation}
\textbf{p} = \textbf{p}_{x}+\textbf{p}_{y}
=\alpha_{\perp} E_{x} \hat{\textbf{x}}+ \alpha_\parallel E_{y}\hat{\textbf{y}}.
\label{eq-p-prop-E-xy-CO2}
\end{equation}
Following \cite{Feynman-Lectures-vol-2}, let us consider $E_x=E_y$ (it is $E_1=E_2$, in Feynman's notation), which means that $\textbf{E}$ is applied at $45\degree$.
Note that, in this case, the induced dipole moment $\textbf{p}$ is not in the same direction as the electric field $\textbf{E}$, as shown in Fig. \ref{fig:molecula-campo-45} (this figure corresponds to Fig. 31-1(a) of Ref. \cite{Feynman-Lectures-vol-2}).
This occurs because $\alpha_{\parallel} \neq \alpha_{\bot}$, which results in $p_y \neq p_x$, even with $E_x=E_y$.
The explanation given by Feynman \cite{Feynman-Lectures-vol-2}, 
in the context of a crystal, can be directly applied to this case of a $\text{CO}_{2}$ molecule:
\begin{quote}
\textit{``The polarization is no longer in the same direction as the electric field. 
	You can see how that might come about. There may be charges which can move easily up and down, 
	but which are rather stiff for sidewise motions. When a force is applied at  $45\degree$, 
	the charges move farther up than they do toward the side. 
	The displacements are not in the direction of the external force, because there are asymmetric internal elastic forces.''}
\end{quote}
Moreover, 
if we replace \textit{``polarization of a crystal''} by \textit{``dipole moment of a $\text{CO}_{2}$ molecule''},
in the text below \cite{Feynman-Lectures-vol-2}, we have a comment also valid for a $\text{CO}_{2}$ molecule:
\begin{quote}
\textit{``There is, of course, nothing special about $45\degree$. It is generally true that the induced polarization of a crystal is not in the direction of the electric field.''}
\end{quote}
This can be seen, for instance, in Fig. \ref{fig:molecula-campo-60} where we consider $\textbf{E}$ applied at $60 \degree$ (this results in $E_x \neq E_y$).

In the most general case, when a field $\textbf{E}=\textbf{E}_{x}+\textbf{E}_{y}+\textbf{E}_{z}$ is applied, we have
\begin{equation}
	\textbf{p} = \alpha_\bot E_{x}\hat{\textbf{x}}+ \alpha_\parallel E_{y}\hat{\textbf{y}}+ \alpha_\bot E_{z}\hat{\textbf{z}}.
	\label{eq-p-prop-E-geral-CO2}
\end{equation}
Note that, differently from the isotropic particle, for which we can always write Eq. \eqref{p-beta-E}, for an anisotropic particle we cannot simply write $\textbf{p}$ as the field $\textbf{E}$ multiplied by a constant, which means that the induced dipole moment $\textbf{p}$ may not be in the same direction as the field $\textbf{E}$.
In this case, we write an equation with a similar structure of Eq. \eqref{p-beta-E}
(something that characterizes the polarizability, multiplied by the electric field)
expressing Eq. \eqref{eq-p-prop-E-geral-CO2} as a matrix equation, i.e.
\begin{equation}
\left[\begin{array}{c}
p_{x}\\
p_{y}\\
p_{z}
\end{array}\right]=\left[\begin{array}{ccc}
\alpha_\bot & 0 & 0\\
0 & \alpha_\parallel & 0\\
0 & 0 & \alpha_\perp
\end{array}\right]\left[\begin{array}{c}
E_{x}\\
E_{y}\\
E_{z}
\end{array}\right]. \label{eq-p-E-matrix-anisotr}
\end{equation}
Writing in a more compact form, one has
\begin{equation}
\textbf{p}= \boldsymbol{\alpha} \textbf{E},
\label{p-beta-E-tensor}
\end{equation}
where, here, the vectors $\textbf{p}$ and $\textbf{E}$ are represented by the column matrices in left and right hand side of Eq. \eqref{eq-p-E-matrix-anisotr}, respectively, whereas 
\begin{equation}
	\boldsymbol{\alpha}=\left[\begin{array}{ccc}
	\alpha_\bot & 0 & 0\\
	0 & \alpha_\parallel & 0\\
	0 & 0 & \alpha_\perp
	\end{array}\right]. \label{eq:alpha-matrix-anisotr}
\end{equation}
Note that, unlike Eq. \eqref{p-beta-E}, in Eq. \eqref{p-beta-E-tensor} we are using the bold symbol
$\boldsymbol{\alpha}$.
Comparing Eq. \eqref{p-beta-E} with \eqref{p-beta-E-tensor}, one can see that
the latter one is more complicated, because
it shows that the vector $\textbf{p}$ is related with the vector $\textbf{E}$ by a second-rank Cartesian tensor, represented by the matrix $\boldsymbol{\alpha}$, 
whereas the former [Eq. \eqref{p-beta-E}] shows that for a spherically symmetric charge distribution
[illustrated in Fig. \ref{fig:atomo1}] the vector $\textbf{p}$ is related to the vector $\textbf{E}$ by a single number (a scalar) $\alpha$ \cite{Bonin-Kresin-1956}.

We can also write Eq. \eqref{p-beta-E-tensor} in index notation (since this notation is commonly used to deal with tensors, one can also call it as tensor notation), which is given by
\begin{equation}
p_i = \sum_{j=1}^3 \alpha_{ij} E_j, \label{eq-p-E-index-isotr-2}
\end{equation}
where $\alpha_{ij}$ are the elements of the matrix $\boldsymbol{\alpha}$, which is the representation in the system $xyz$ of the polarizability tensor $\overleftrightarrow{\alpha}$.
%
%
According to Feynman \cite{Feynman-Lectures-vol-2},
\begin{quote}
	\textit{``The tensor $\alpha_{ij}$ should really be called a ``tensor of second rank,'' because it has two indexes. A vector - with one index - is a tensor of the first rank, and a scalar - with no index - is a tensor of zero rank.''}
\end{quote}
It is important to remark that $\overleftrightarrow{\alpha}$ is represented in $xyz$ system by a diagonal matrix in Eq. \eqref{eq-p-E-matrix-anisotr}, which occurs because the $\text{CO}_2$ molecule was chosen having its principal axes parallel to the $xyz$ axes. However, $\overleftrightarrow{\alpha}$ can have a non-diagonal representation, as discussed in Secs. \ref{rotated-molecule-matrix-representation} and \ref{rotated-system}.
%

\subsection{A visual representation of the $\text{CO}_{2}$ polarizability tensor}
\label{visulization}

The polarizability tensor $\overleftrightarrow{\alpha}$ establishes the connection between the induced dipole moment $\textbf{p}$
and the incident field $\textbf{E}$.
Thus, a way to have a certain visual representation of $\overleftrightarrow{\alpha}$, is by means of a visual representation of the behavior of $\textbf{p}$ in terms of $\textbf{E}$. 

Let us consider the $\text{CO}_{2}$ molecule illustrated in Fig. \ref{fig:molecula-eixo-y}, and apply $\textbf{E}$ in different directions, but with a same magnitude. 
In a first moment, we also consider, for simplicity, $\textbf{E}$ having only $x$- or $y$-component.
In other words, we consider the situations as illustrated in Figs. \ref{fig:molecula-campo-x}
and \ref{fig:molecula-campo-y}, 
but now considering that all applied fields have the same magnitude.

\begin{figure}[h]
\centering  
\subfigure[]{\label{fig:co2-campo-x1}\epsfig{file=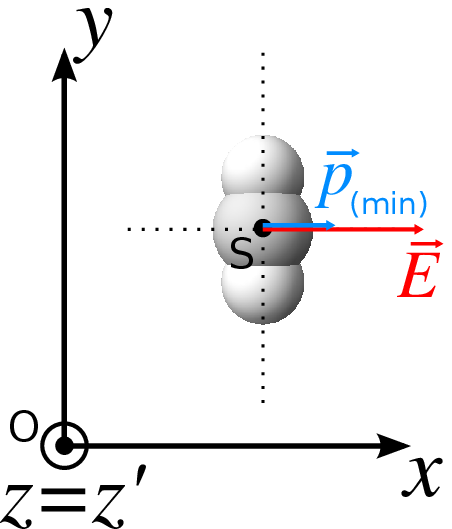,width=0.45 \linewidth}}
\hspace{5mm}
\subfigure[]{\label{fig:co2-campo-x2}\epsfig{file=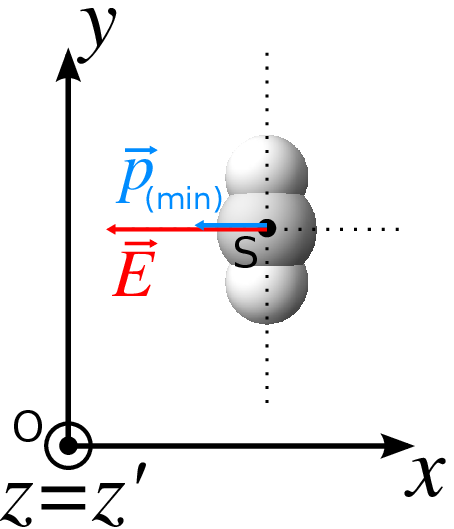,width=0.45 \linewidth}}
\caption{\footnotesize{
A $\text{CO}_{2}$ molecule, oriented in space in a such way that its principal axes 
(dotted lines) are parallel to those  of the laboratory $xyz$-system.
Note that the $z$-axis appears perpendicular to the paper.
Here it is illustrated the two possible induced dipole moments $\textbf{p}$, in the presence of fields $\textbf{E}$ with a same magnitude, pointing along $y$-direction. 
(a) The induced dipole moment $\textbf{p}_{\text{(min)}}$, 
when $\textbf{E}$ points in the positive $x$-direction.
(b) The induced dipole moment $\textbf{p}_{\text{(min)}}$, 
when $\textbf{E}$ points in the negative $x$-direction.
}}
\label{co2-campo-x}
\end{figure}

\begin{figure}[h]
\centering
\subfigure[]{\label{fig:co2-campo-y1}\epsfig{file=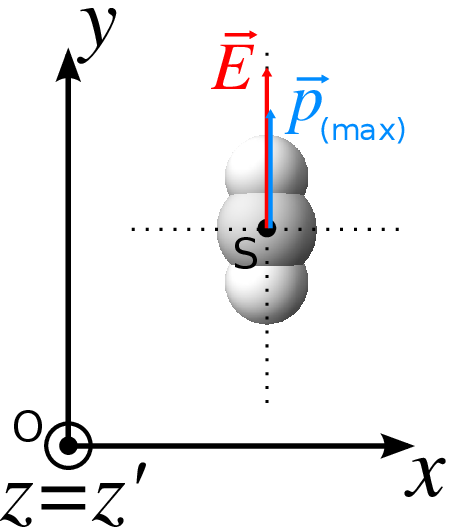,width=0.45 \linewidth}}
\hspace{5mm}
\subfigure[]{\label{fig:co2-campo-y2}\epsfig{file=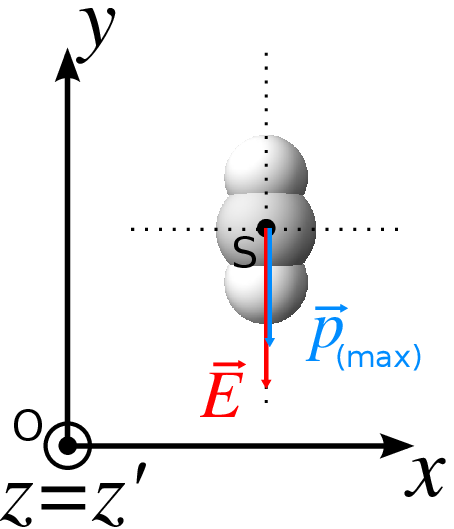,width=0.45 \linewidth}}
\caption{\footnotesize{
A $\text{CO}_{2}$ molecule, oriented in space in a such way that its principal axes 
(dotted lines) are parallel to those  of the laboratory $xyz$-system.
Note that the $z$-axis appears perpendicular to the paper.
Here it is illustrated the two possible induced dipole moments $\textbf{p}$, in the presence of fields $\textbf{E}$ with a same magnitude, pointing along the $y$-direction. 
(a) The induced dipole moment $\textbf{p}_{\text{(max)}}$, 
when $\textbf{E}$ points in the positive $y$-direction.
(b) The induced dipole moment $\textbf{p}_{\text{(max)}}$, 
when $\textbf{E}$ points in the negative $y$-direction.
}}
\label{co2-campo-y}
\end{figure}

\begin{figure}[h]
\centering
\epsfig{file=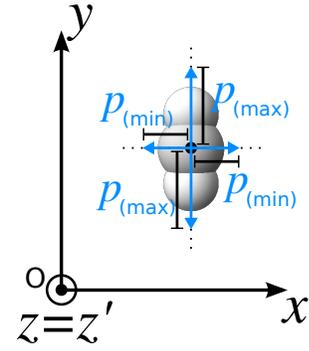,width=0.45 \linewidth}
\caption{\footnotesize{
A superposition of the images taken from Figs. \ref{co2-campo-x} and \ref{co2-campo-y}.
Here it is illustrated the four possible induced dipole moments $\textbf{p}$ for the $\text{CO}_{2}$ molecule shown in Fig. \ref{fig:molecula-eixo-y}, in the presence of fields $\textbf{E}$ with a same magnitude, pointing along $x$- or $y$-direction. 
}}
\label{co2-germinal}
\end{figure}

When $\textbf{E}$ points in the $x$-direction, one has the dipole moment, renamed
$\textbf{p}_{\text{(min)}}$ (this nomenclature is explained later), given by $\textbf{p}_{\text{(min)}}=\alpha_{\bot}\textbf{E}$ [see Eq. \eqref{eq-p-prop-E-perp-CO2}], as illustrated in Fig. \ref{co2-campo-x}. 
If $\textbf{E}$ points in the $y$-direction, one has the dipole moment, renamed
$\textbf{p}_{\text{(max)}}$, given by $\textbf{p}_{\text{(max)}}=\alpha_{\parallel}\textbf{E}$ [see Eq. \eqref{eq-p-prop-E-axis-CO2}], as illustrated in Fig. \ref{co2-campo-y}. 
When making a superposition of the images taken from Figs. \ref{co2-campo-x} and \ref{co2-campo-y}, we have Fig. \ref{co2-germinal}, which illustrates the behavior of $\textbf{p}$ in terms of $\textbf{E}$ (with this field having the same magnitude in all the cases, and pointing to $x$- or $y$-direction). 
We can say that Fig. \ref{co2-germinal} is a germinal visual representation of $\overleftrightarrow{\alpha}$.

\begin{figure}[h]
\centering
\epsfig{file=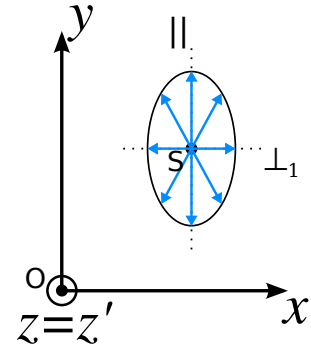,width=0.45 \linewidth}
\caption{\footnotesize{
Ellipse corresponding to Eq. \eqref{elipse-2d},
which defines all possible induced dipole moments $\textbf{p}=\textbf{p}_x+\textbf{p}_y$ (which go from the the point $S$ to the curve of the ellipse) for the $\text{CO}_{2}$ molecule shown in Fig. \ref{fig:molecula-eixo-y}, in the presence of fields $\textbf{E}=\textbf{E}_x+\textbf{E}_y$ with a same magnitude.
This figure is an extension of Fig. \ref{co2-germinal}.	
}}
\label{elipse}
\end{figure}

The visual representation of $\overleftrightarrow{\alpha}$ in Fig. \ref{co2-germinal} does not take into account the situation in which $\textbf{E}=\textbf{E}_x+\textbf{E}_y$. 
For this more general case, one has $\textbf{p}$ given by Eq. \eqref{eq-p-prop-E-geral-CO2} (with $E_z=0$),
and two particular cases illustrated in Figs. \ref{fig:molecula-campo-45} and \ref{fig:molecula-campo-60}.
Then, let us consider again the application, on a $\text{CO}_{2}$ molecule as illustrated in Fig. \ref{fig:molecula-eixo-y}, but with $\textbf{E}=\textbf{E}_x+\textbf{E}_y$ applied in different directions, under the condition that all the applied fields have a same magnitude
\begin{equation}
	E^{2}=E_{x}^{2}+E_{y}^{2}.
\end{equation}
Using Eqs. \eqref{P1-CO2} and \eqref{P2-CO2}, we get
\begin{equation}
\frac{p_{x}^{2}}{E^{2}\alpha_{\perp}^{2}}+\frac{p_{y}^{2}}{E^{2}\alpha_{\parallel}^{2}}=1.
\label{pre-elipse}
\end{equation}
We define:
\begin{eqnarray}
	p_{\text{(min)}}^{2}=E^{2}\alpha_{\perp}^{2},
		\label{p-min-def}
	\\
	p_{\text{(max)}}^{2}=E^{2}\alpha_{\parallel}^{2}.
	\label{p-max-def}
\end{eqnarray}
Then, we can write \eqref{pre-elipse} as
\begin{equation}
\frac{p_{x}^{2}}{p_{\text{(\text{min)}}}^{2}}+\frac{p_{y}^{2}}{p_{\text{(\text{max)}}}^{2}}=1.
\label{elipse-2d}
\end{equation}
Note that this is the equation of an ellipse centered at the point $S$ (since we are considering the vectors $\textbf{E}$ and $\textbf{p}$ with their origin in this point), as shown in Fig. \ref{elipse}. 
The values of $p_{x}$ and $p_{y}$ satisfying Eq. \eqref{elipse-2d} define all possible induced dipole moments $\textbf{p}$ for the $\text{CO}_{2}$ molecule shown in Fig. \ref{fig:molecula-eixo-y}, in the presence of fields $\textbf{E}=\textbf{E}_x+\textbf{E}_y$ with a same magnitude $E$.
Note that the minimum magnitude of $\textbf{p}$ is $p_{\text{(min)}}$, along the minor axis of the ellipse, and the maximum magnitude is $p_{\text{(max)}}$, along the larger axis. This justifies the nomenclature introduced above.

\begin{figure}[h]
\centering
\epsfig{file=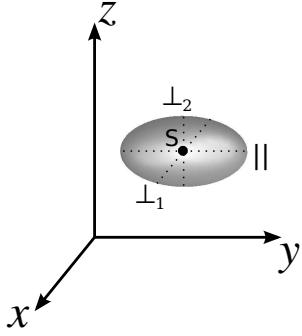,width=0.45 \linewidth}
\caption{\footnotesize{
Lamé's ellipsoid, corresponding to Eq. \eqref{elipsoide-3d},
which defines all possible induced dipole moments $\textbf{p}=\textbf{p}_x+\textbf{p}_y+\textbf{p}_z$ for the $\text{CO}_{2}$ molecule shown in Fig. \ref{fig:molecula-eixo-y}, in the presence of fields $\textbf{E}=\textbf{E}_x+\textbf{E}_y+\textbf{E}_z$ with a same magnitude.
In this sense, this Lamé ellipsoid is a way to get some visualization of the polarizability tensor
for the $\text{CO}_{2}$ molecule shown in Fig. \ref{fig:molecula-eixo-y}.
This figure is an extension of Fig. \ref{elipse}.			%
}}
\label{elipsoide}
\end{figure}

For a more general visual representation of $\overleftrightarrow{\alpha}$, let us consider the application, on a $\text{CO}_{2}$ molecule as illustrated in Fig. \ref{fig:molecula-eixo-y}, of a field $\textbf{E}=\textbf{E}_x+\textbf{E}_y+
\textbf{E}_z$ applied in different directions, under the condition that 
all the applied fields have a same magnitude
\begin{equation}
	E^{2}=E_{x}^{2}+E_{y}^{2}+E_{z}^{2}.
	\label{eq:ellipsoid-passo-0}
\end{equation}
Using Eqs. \eqref{P1-CO2} - \eqref{P3-CO2}, we have
\begin{equation}
	\frac{p_{x}^{2}}{E^{2}\alpha_{\perp}^{2}}+\frac{p_{y}^{2}}{E^{2}\alpha_{\parallel}^{2}}+	\frac{p_{z}^{2}}{E^{2}\alpha_{\perp}^{2}}=1.
	\label{pre-elipsoide}
\end{equation}
Using Eqs. \eqref{p-min-def} and \eqref{p-max-def}, we have
\begin{equation}
	\frac{p_{x}^{2}}{p_{\text{(\text{min)}}}^{2}}
	+\frac{p_{y}^{2}}{p_{\text{(\text{max)}}}^{2}}
	+\frac{p_{z}^{2}}{p_{\text{(\text{min)}}}^{2}}=1,
	\label{elipsoide-3d}
\end{equation}
which is the equation of an ellipsoid (called in this work Lamé's ellipsoid), centered in the point $S$, as shown in Fig. \ref{elipsoide}. 
This Lamé ellipsoid defines all possible induced dipole moments $\textbf{p}$ for the $\text{CO}_{2}$ molecule shown in Fig. \ref{fig:molecula-eixo-y}, in the presence of fields $\textbf{E}$ with a same magnitude.
Since the tensor $\overleftrightarrow{\alpha}$ establishes the behavior of $\textbf{p}$ in terms of $\textbf{E}$, the Lamé ellipsoid is a way to get some visualization of this tensor.
It is important to remark that the Lamé ellipsoid was considered originally for a visual representation of the stress tensor \cite{Fung-1965}. 
Here, we constructed a correspondent Lamé ellipsoid for the polarizability tensor, and simply called it as the Lamé ellipsoid.

\subsection{A visual representation of the polarizability tensor for a rotated $\text{CO}_{2}$ molecule}
\label{rotated-molecule-visual}

In this section, we study what happens if the $\text{CO}_{2}$ molecule, instead of being oriented in space as shown in Fig. \ref{fig:molecula-eixo-y}, is rotated of an angle $\phi$ with the $y$-direction, as illustrated in Fig. \ref{fig:molecula-rotacao}.
First, it is important to consider a preliminary example of a rotating object characterized
by a vector.
Thus, note that the dipole moment vector $\textbf{p}$ of a $\text{H}_2\text{O}$ molecule
changes to a different vector $\tilde{\textbf{p}}$ when this molecule is rotated.
%
Here, in a similar way, the polarizability tensor $\overleftrightarrow{\alpha}$, for the $\text{CO}_{2}$ molecule in Fig. \ref{fig:molecula-eixo-y}, changes to a different tensor $\widetilde{\overleftrightarrow{\alpha}}$ when this molecule is rotated as in Fig. \ref{fig:molecula-rotacao}.
This difference occurs because the polarizability tensor establishes the spatial connection between the induced dipole moment and an incident electric field, thus, when the molecule rotates, this spatial connection changes, which means a change in its polarizability tensor.

As we can see in Fig. \ref{fig:molecula-rotacao}, when the molecule is rotated in the $xy$-plane, and makes an angle $\phi$ with the $y$-direction, its principal axes $\tilde{\parallel}$ and $\tilde{\bot}_1$ are no longer parallel to the $y$ and $x$, respectively.
%
%
As has been done before, let us apply electric fields $\textbf{E}$ on this rotated $\text{CO}_{2}$ molecule, with $\textbf{E}$ applied in different directions, but under the condition that they have a same magnitude $E$.
We could write $\textbf{E}=\textbf{E}_x+\textbf{E}_y+\textbf{E}_z$, as done before, but for convenience
we write
\begin{equation}
	\textbf{E}=\textbf{E}_{\tilde{\bot}_1}+\textbf{E}_{\tilde{\parallel}}+\textbf{E}_{\tilde{\bot}_2},
\end{equation}
where now we are decomposing the field in the directions ${\tilde{\bot}_1}$,
${\tilde{\parallel}}$, and  ${\tilde{\bot}_2}$.
When $\textbf{E}$ points in the  ${\tilde{\bot}_1}$-direction ($\textbf{E}=\textbf{E}_{\tilde{\bot}_1}$), 
one has the dipole moment, renamed $\tilde{\textbf{p}}_{\bot_1}$, given by [see Eq. \eqref{eq-p-prop-E-perp-CO2}]
\begin{equation}
	\tilde{\textbf{p}}_{\bot_1}=\alpha_{\bot}\textbf{E}_{\tilde{\bot}_1},
	\label{p1-min}
\end{equation}
and illustrated in Fig. \ref{co2-campo-x}.
%
If $\textbf{E}$ points in the ${\tilde{\parallel}}$-direction ($\textbf{E}=\textbf{E}_{\tilde{\parallel}}$), 
one has the dipole moment, renamed
$\tilde{\textbf{p}}_{{\tilde{\parallel}}}$, given by [see Eq. \eqref{eq-p-prop-E-axis-CO2}]
\begin{equation}
	\tilde{\textbf{p}}_{\tilde{\parallel}}=\alpha_{\parallel}\textbf{E}_{\tilde{\parallel}},
	\label{p-max}
\end{equation}
as illustrated in Fig. \ref{co2-campo-y}. 
%
When $\textbf{E}$ points in the  ${\tilde{\bot}_2}$-direction ($\textbf{E}=\textbf{E}_{\tilde{\bot}_2}$), 
one has the dipole moment, renamed
$\tilde{\textbf{p}}_{\tilde{\bot}_2}$, given by [see Eq. \eqref{eq-p-prop-E-perp-CO2}]
\begin{equation}
	\tilde{\textbf{p}}_{\tilde{\bot}_2}=\alpha_{\bot}\textbf{E}_{\tilde{\bot}_2}.
	\label{p2-min}
\end{equation}
Note that the constants $\alpha_{\bot}$ and $\alpha_{\parallel}$ appearing in Eqs.
\eqref{p1-min} - \eqref{p2-min} are the same as those in Eqs.
\eqref{eq-p-prop-E-axis-CO2} and \eqref{eq-p-prop-E-perp-CO2}.
One can write
\begin{equation}
		E^{2}={E}_{\tilde{\bot}_1}^2+{E}_{\tilde{\parallel}}^2+{E}_{\tilde{\bot}_2}^2.
		\label{E2-tilde}
\end{equation}
Using Eqs. \eqref{p1-min}-\eqref{p2-min} in Eq. \eqref{E2-tilde}, we have
\begin{equation}
	\frac{\tilde{{p}}_{\tilde{\bot}_1}}{E^{2}\alpha_{\perp}^{2}}
	+\frac{\tilde{{p}}_{\tilde{\parallel}}}{E^{2}\alpha_{\parallel}^{2}}
	+\frac{\tilde{{p}}_{\tilde{\bot}_2}}{E^{2}\alpha_{\perp}^{2}}=1.
	\label{pre-elipsoide-rotated}
\end{equation}
In this way, from Eqs. \eqref{p-min-def} and \eqref{p-max-def}, we have
%
\begin{equation}
	\frac{\tilde{{p}}_{\tilde{\bot}_1}}{p_{\text{\text{(min)}}}^{2}}
	+\frac{\tilde{{p}}_{\tilde{\parallel}}}{p_{\text{\text{(max)}}}^{2}}
	+\frac{\tilde{{p}}_{\tilde{\bot}_2}}{p_{\text{\text{(min)}}}^{2}}=1.
	\label{elipsoide-3d-rotated}
\end{equation}
This is the equation of the Lamé ellipsoid (centered in the point $S$), 
which is a visual representation of the polarizability tensor
$\widetilde{\overleftrightarrow{\alpha}}$, for the rotated $\text{CO}_{2}$ molecule in Fig. \ref{fig:molecula-rotacao}. 
As we can see, as the $\text{CO}_{2}$ molecule rotates, this Lamé ellipsoide rotates together.

\subsection{Matrix representation of the polarizability tensor for a rotated $\text{CO}_{2}$ molecule}
\label{rotated-molecule-matrix-representation}

\begin{figure}[h]
\centering  
\subfigure[]{\label{molecula-rotacao-3d}\epsfig{file=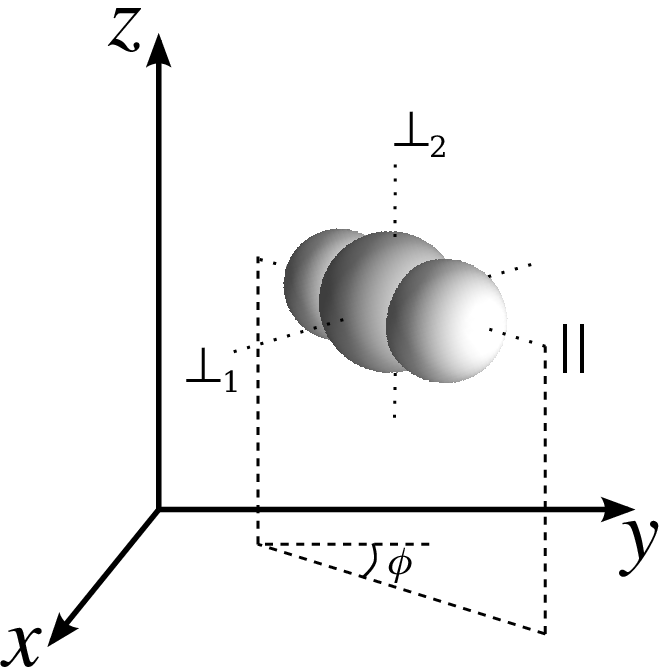, width=0.45 \linewidth}}
\hspace{2mm}
\subfigure[]{\label{molecula-rotacao}\epsfig{file=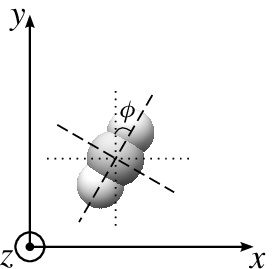, width=0.45 \linewidth}}
\caption{\footnotesize{ 
		Illustration of the rotation  by an angle $\phi$ 
		(with the $y$-direction)
		of the $\text{CO}_{2}$ molecule illustrated in Fig. \ref{molecula-eixo-y}. 
		In (a), we show a 3D, and in (b), a 2D visualization. 
}}
\label{fig:molecula-rotacao}
\end{figure}

\begin{figure}[h]
	\centering  
	\subfigure[]{\label{fig:molecula-rot-campo-x}\epsfig{file=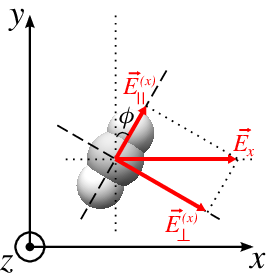, width=0.45 \linewidth}}
	\hspace{2mm}
	\subfigure[]{\label{fig:molecula-rot-campo-x-pol}\epsfig{file=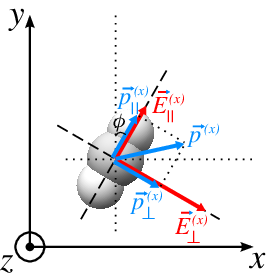, width=0.45 \linewidth}}
	\caption{\footnotesize{ 
			(a) $\textbf{E}_x$ and its components in $\parallel$ and $\perp$ directions with respect to the molecule axis.
			(b) The correspondent polarization $\textbf{p}^{(x)}$ and its components in $\parallel$ and $\perp$ directions.
	}}
\end{figure}

In this section, we discuss the representation of $\widetilde{\overleftrightarrow{\alpha}}$ (corresponding
to the $\text{CO}_{2}$ molecule illustrated in Fig. \ref{fig:molecula-rotacao}) in the $xyz$-system.
If we apply an electric field $\textbf{E}_{x}$ in the $x$-direction, one can obtain that this field can be decomposed into two fields, namely $\textbf{E}_{\parallel}^{\left(x\right)}$ and $\textbf{E}_{\perp}^{\left(x\right)}$ [the superscript $(x)$ indicates that these fields result from the application of $\textbf{E}_{x}$], which are applied in the directions parallel and perpendicular to the molecule axis, respectively [see Fig. \ref{fig:molecula-rot-campo-x}]. 
We can write their magnitudes in terms of $|\textbf{E}_{x}|=E_{x}$ as
\begin{align}
E_{\parallel}^{\left(x\right)} & =E_{x}\sin\phi,\\
E_{\perp}^{\left(x\right)} & =E_{x}\cos\phi,
\end{align}
so that we can write $\textbf{E}_{\parallel}^{\left(x\right)}$ as
\begin{align}
\textbf{E}_{\parallel}^{\left(x\right)} & =E_{\parallel}^{\left(x\right)}\sin\phi\hat{\textbf{x}}+E_{\parallel}^{\left(x\right)}\cos\phi\hat{\textbf{y}} \nonumber \\
& =E_{x}\sin^{2}\phi\hat{\textbf{x}}+E_{x}\cos\phi\sin\phi\hat{\textbf{y}}, \label{eq-E-paral-x}
\end{align}
and $\textbf{E}_{\perp}^{\left(x\right)}$ as
\begin{align}
\textbf{E}_{\perp}^{\left(x\right)} & =E_{\perp}^{\left(x\right)}\cos\phi\hat{\textbf{x}}-E_{\perp}^{\left(x\right)}\sin\phi\hat{\textbf{y}} \nonumber \\
& =E_{x}\cos^{2}\phi\hat{\textbf{x}}-E_{x}\cos\phi\sin\phi\hat{\textbf{y}}. \label{eq-E-perp-x}
\end{align}
Thus, according to Eqs. \eqref{eq-p-prop-E-axis-CO2} and \eqref{eq-E-paral-x}, the field $\textbf{E}_{\parallel}^{\left(x\right)}$ produces a dipole moment $\textbf{p}_{\parallel}^{\left(x\right)}$, given by
\begin{align}
\textbf{p}_{\parallel}^{\left(x\right)} & =\alpha_{\parallel}\textbf{E}_{\parallel}^{\left(x\right)}\nonumber \\
& =\alpha_{\parallel}E_{x}\sin^{2}\phi\hat{\textbf{x}}+\alpha_{\parallel}E_{x}\cos\phi\sin\phi\hat{\textbf{y}}, \label{eq-p-paral-x}
\end{align}
whereas, according to Eqs. \eqref{eq-p-prop-E-perp-CO2} and \eqref{eq-E-perp-x}, the field $\textbf{E}_{\perp}^{\left(x\right)}$ produces a dipole moment $\textbf{p}_{\perp}^{\left(x\right)}$, given by
\begin{align}
\textbf{p}_{\perp}^{\left(x\right)} & =\alpha_{\perp}\textbf{E}_{\perp}^{\left(x\right)}\nonumber \\
& =\alpha_{\perp}E_{x}\cos^{2}\phi\hat{\textbf{x}}-\alpha_{\perp}E_{x}\cos\phi\sin\phi\hat{\textbf{y}}. \label{eq-p-perp-x}
\end{align}
The dipole moment vectors $\textbf{p}_{\parallel}^{\left(x\right)}$ and $\textbf{p}_{\perp}^{\left(x\right)}$ produce a resultant dipole moment vector given by the sum $\textbf{p}^{\left(x\right)}=\textbf{p}_{\parallel}^{\left(x\right)}+\textbf{p}_{\perp}^{\left(x\right)}$ [see Fig. \ref{fig:molecula-rot-campo-x-pol}], so that, from Eqs. \eqref{eq-p-paral-x} and \eqref{eq-p-perp-x}, it can be written as
\begin{align}
\textbf{p}^{\left(x\right)} & =(\alpha_{\parallel}\sin^{2}\phi+\alpha_{\perp}\cos^{2}\phi)E_{x}\hat{\textbf{x}}\nonumber \\
& \quad+(\alpha_{\parallel}-\alpha_{\perp})\sin\phi\cos\phi E_{x}\hat{\textbf{y}}. \label{eq-p-x}
\end{align}

\begin{figure}[h]
	\centering  
	\subfigure[]{\label{fig:molecula-rot-campo-y}\epsfig{file=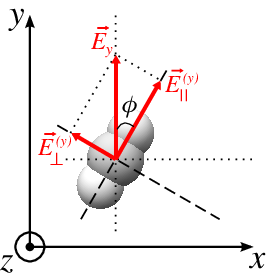, width=0.45 \linewidth}}
	\hspace{2mm}
	\subfigure[]{\label{fig:molecula-rot-campo-y-pol}\epsfig{file=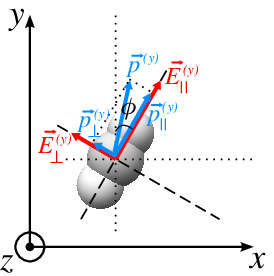, width=0.45 \linewidth}}
	\caption{\footnotesize{ 
			(a) $\textbf{E}_y$ and its components in $\parallel$ and $\perp$ directions with respect to the molecule axis.
			(b) The correspondent polarization $\textbf{p}^{(y)}$ and its components in $\parallel$ and $\perp$ directions.
	}}
\end{figure}

If we apply an electric field $\textbf{E}_{y}$ in the $y$-direction, one can obtain that this field can be decomposed into two fields, namely $\textbf{E}_{\parallel}^{\left(y\right)}$ and $\textbf{E}_{\perp}^{\left(y\right)}$ [the superscript $(y)$ indicates that these fields result from the application of $\textbf{E}_{y}$], which are applied in the directions parallel and perpendicular to the molecule axis, respectively [see Fig. \ref{fig:molecula-rot-campo-y}]. 
We can write their magnitudes in terms of $|\textbf{E}_{y}|=E_{y}$ as
\begin{align}
E_{\parallel}^{\left(y\right)} & =E_{y}\cos\phi,\\
E_{\perp}^{\left(y\right)} & =E_{y}\sin\phi,
\end{align}
so that we can write $\textbf{E}_{\parallel}^{\left(y\right)}$ as
\begin{align}
\textbf{E}_{\parallel}^{\left(y\right)} & =E_{\parallel}^{\left(y\right)}\sin\phi\hat{\textbf{x}}+E_{\parallel}^{\left(y\right)}\cos\phi\hat{\textbf{y}}\nonumber \\
& =E_{y}\sin\phi\cos\phi\hat{\textbf{x}}+E_{y}\cos^{2}\phi\hat{\textbf{y}}, \label{eq-E-paral-y}
\end{align}
and $\textbf{E}_{\perp}^{\left(y\right)}$ as
\begin{align}
\textbf{E}_{\perp}^{\left(y\right)} & =-E_{\perp}^{\left(y\right)}\cos\phi\hat{\textbf{x}}+E_{\perp}^{\left(y\right)}\sin\phi\hat{\textbf{y}} \nonumber \\
& =-E_{y}\sin\phi\cos\phi\hat{\textbf{x}}+E_{y}\sin^{2}\phi\hat{\textbf{y}}. \label{eq-E-perp-y}
\end{align}
Thus, according to Eqs. \eqref{eq-p-prop-E-axis-CO2} and \eqref{eq-E-paral-y}, the field $\textbf{E}_{\parallel}^{\left(y\right)}$ produces a dipole moment $\textbf{p}_{\parallel}^{\left(y\right)}$, given by
\begin{align}
\textbf{p}_{\parallel}^{\left(y\right)} & =\alpha_{\parallel}\textbf{E}_{\parallel}^{\left(y\right)}\nonumber \\
& =\alpha_{\parallel}E_{y}\sin\phi\cos\phi\hat{\textbf{x}}+\alpha_{\parallel}E_{y}\cos^{2}\phi\hat{\textbf{y}}, \label{eq-p-paral-y}
\end{align}
whereas, according to Eqs. \eqref{eq-p-prop-E-perp-CO2} and \eqref{eq-E-perp-y}, the field $\textbf{E}_{\perp}^{\left(y\right)}$ produces a dipole moment $\textbf{p}_{\perp}^{\left(y\right)}$, given by
\begin{align}
\textbf{p}_{\perp}^{\left(y\right)} & =\alpha_{\perp}\textbf{E}_{\perp}^{\left(y\right)}\nonumber \\
& =-\alpha_{\perp}E_{y}\sin\phi\cos\phi\hat{\textbf{x}}+\alpha_{\perp}E_{y}\sin^{2}\phi\hat{\textbf{y}}. \label{eq-p-perp-y}
\end{align}
The dipole moment vectors $\textbf{p}_{\parallel}^{\left(y\right)}$ and $\textbf{p}_{\perp}^{\left(y\right)}$ produce a resultant dipole moment vector given by the sum $\textbf{p}^{\left(y\right)}=\textbf{p}_{\parallel}^{\left(y\right)}+\textbf{p}_{\perp}^{\left(y\right)}$ [see Fig. \ref{fig:molecula-rot-campo-y-pol}], so that, from Eqs. \eqref{eq-p-paral-y} and \eqref{eq-p-perp-y}, it can be written as
\begin{align}
\textbf{p}^{\left(y\right)} & =\left(\alpha_{\parallel}-\alpha_{\perp}\right)\sin\phi\cos\phi E_{y}\hat{\textbf{x}}\nonumber \\
& \quad+\left(\alpha_{\parallel}\cos^{2}\phi+\alpha_{\perp}\sin^{2}\phi\right)E_{y}\hat{\textbf{y}}. \label{eq-p-y}
\end{align}

We can also apply an electric field $\textbf{E}_{z}$ in the $z$-direction, but in this case, we simply have that this field, according to Eq. \eqref{eq-p-prop-E-perp-CO2}, produces a dipole moment $\textbf{p}^{\left(z\right)}$, given by
\begin{equation}
\textbf{p}^{\left(z\right)} =\alpha_{\perp}\textbf{E}_{z}=\alpha_{\perp} E_{z} \hat{\textbf{z}}, \label{eq-p-z}
\end{equation}
since the molecule still has a principal axis (perpendicular to the molecule axis) parallel to the $z$-direction.

\begin{figure}[h]
\centering
		\epsfig{file=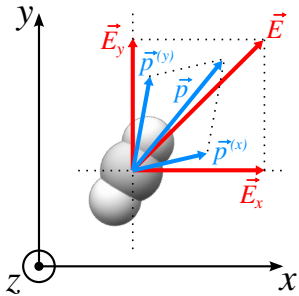,width=0.45 \linewidth}
		\caption{\footnotesize{
				An illustration of the relation between  $\textbf{p}$ and $\textbf{E}$, given in Eq. \eqref{p-torto}, for the case $E_x=E_y$, which means that $\textbf{E}$ is applied at $45\degree$.
		}}
		\label{fig:molecula-rot-campo-xy}
\end{figure}

As we did in the previous section, let us investigate a superposition of the fields $\textbf{E}_{x}$ and $\textbf{E}_{y}$.
Thus, if we apply an electric field $\textbf{E}=E_{x}\hat{\textbf{x}}+E_{y}\hat{\textbf{y}}$, it produces a dipole moment $\textbf{p}$ that can be written as a sum of the dipole moments produced by the fields $\textbf{E}_x$ and $\textbf{E}_y$ separately.
Thus, we can write $\textbf{p}$ as
\begin{equation}
\textbf{p}=\textbf{p}^{\left(x\right)}+\textbf{p}^{\left(y\right)},
\end{equation}
which, from Eqs. \eqref{eq-p-x} and \eqref{eq-p-y}, can be written as
\begin{align}
\textbf{p} = &  [(\alpha_{\parallel}\sin^{2}\phi+\alpha_{\perp}\cos^{2}\phi)E_{x}+(\alpha_{\parallel}-\alpha_{\perp})\sin\phi
\nonumber \\
&\times \cos\phi E_{y}]\hat{\textbf{x}}+ [(\alpha_{\parallel}-\alpha_{\perp})\sin\phi\cos\phi E_{x} \nonumber \\
& +(\alpha_{\parallel}\cos^{2}\phi+\alpha_{\perp}\sin^{2}\phi)E_{y}]\hat{\textbf{y}}.
\label{p-torto}
\end{align}
An illustration of the relation between  $\textbf{p}$ and $\textbf{E}$, given in Eq. \eqref{p-torto}, is shown in Fig. \ref{fig:molecula-rot-campo-xy}, for the case $E_x=E_y$, which means that $\textbf{E}$ is applied at $45\degree$.
Figure \ref{fig:molecula-rot-campo-xy} corresponds to Fig. 31-1(b) of Ref. \cite{Feynman-Lectures-vol-2}. 
When a general field $\textbf{E} = E_{x} \hat{\textbf{x}} + E_{y} \hat{\textbf{y}} + E_{z} \hat{\textbf{z}}$ is applied, we have
\begin{equation}
\textbf{p}=\textbf{p}^{\left(x\right)}+\textbf{p}^{\left(y\right)}+\textbf{p}^{\left(z\right)},
\end{equation}
which, from Eqs. \eqref{eq-p-x}, \eqref{eq-p-y} and \eqref{eq-p-z}, can be written as
\begin{align}
\textbf{p} = &  [(\alpha_{\parallel}\sin^{2}\phi+\alpha_{\perp}\cos^{2}\phi)E_{x}+(\alpha_{\parallel}-\alpha_{\perp})\sin\phi
\nonumber \\
&\times \cos\phi E_{y}]\hat{\textbf{x}}+ [(\alpha_{\parallel}-\alpha_{\perp})\sin\phi\cos\phi E_{x} \nonumber \\
& +(\alpha_{\parallel}\cos^{2}\phi+\alpha_{\perp}\sin^{2}\phi)E_{y}]\hat{\textbf{y}}+\alpha_{\perp}E_{z}\hat{\textbf{z}}.
\end{align}
We can express this equation as a matrix equation as 
\begin{equation}
	\textbf{p}= \widetilde{\boldsymbol{\alpha}} \textbf{E}
	\label{p-beta-tilde-E-tensor}
\end{equation}
where, here, $\textbf{p}$ and $\textbf{E}$ are represented by the column matrices, and $\tilde{\boldsymbol{\alpha}}$ 
is given by
\begin{equation}
\widetilde{\boldsymbol{\alpha}}=\left[\begin{array}{ccc}
\alpha_{\parallel}\sin^{2}\phi+\alpha_{\perp}\cos^{2}\phi & (\alpha_{\parallel}-\alpha_{\perp})\sin\phi\cos\phi & 0\\
(\alpha_{\parallel}-\alpha_{\perp})\sin\phi\cos\phi & \alpha_{\parallel}\cos^{2}\phi+\alpha_{\perp}\sin^{2}\phi & 0\\
0 & 0 & \alpha_{\perp}
\end{array}\right], \label{eq-beta-anisotr-rot}
\end{equation}
which is the representation in the system $xyz$ of the
polarizability tensor $\widetilde{\overleftrightarrow{\alpha}}$ of
the rotated $\text{CO}_{2}$ molecule illustrated in Fig. \ref{fig:molecula-rotacao}.
Comparing the matrix $\boldsymbol{\alpha}$ in
Eq. \eqref{eq-p-E-matrix-anisotr}, 
with the matrix $\widetilde{\boldsymbol{\alpha}}$ in Eq. \eqref{eq-beta-anisotr-rot}, one can see that in the latter appears off-diagonal elements.
Furthermore, note that $\widetilde{\alpha}_{xy}=\widetilde{\alpha}_{yx}=(\alpha_{\parallel}-\alpha_{\perp})\sin\phi\cos\phi$, $\widetilde{\alpha}_{xz}=\widetilde{\alpha}_{zx}=0$ and $\widetilde{\alpha}_{yz}=\widetilde{\alpha}_{zy}=0$.
In a compact form,
\begin{equation}
	\tilde{\alpha}_{ij} = \tilde{\alpha}_{ji},
	\label{eq-symmetry-tilde}
\end{equation}
which means that the matrix $\widetilde{\boldsymbol{\alpha}}$ is symmetric.

\begin{figure}[h]
\centering
		\epsfig{file=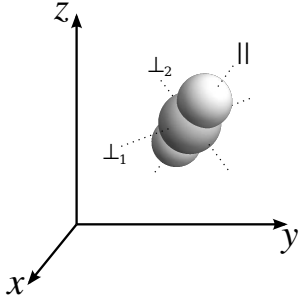,width=0.45 \linewidth}
		\caption{\footnotesize{ 
				Illustration of a $\text{CO}_{2}$ molecule with an arbitrary orientation with respect 
				to the coordinate axes $xyz$.
			}}
		\label{fig:molecula-rot-geral}
\end{figure}

Following Feynman \cite{Feynman-Lectures-vol-2},
\begin{quote}
\textit{``We want now to treat the general case of an arbitrary orientation of a crystal with respect to the coordinate axes.''}
\end{quote}
Here, we replace, in this Feynman sentence, a crystal by a $\text{CO}_{2}$ molecule, so that
we are going to discuss the case of a $\text{CO}_{2}$ molecule with an arbitrary orientation with respect 
to the coordinate axes $xyz$, as illustrated in Fig. \ref{fig:molecula-rot-geral}.
In this case,  the components of $\textbf{p}$ are related with the components of $\textbf{E}$ 
by \cite{Feynman-Lectures-vol-2,Griffiths-Electrodynamics-1999}
\begin{equation}
\begin{array}{l}
p_{x}=\alpha_{xx}E_{x}+\alpha_{xy}E_{y}+\alpha_{xz}E_{z},\\
p_{y}=\alpha_{yx}E_{x}+\alpha_{yy}E_{y}+\alpha_{yz}E_{z},\\
p_{z}=\alpha_{zx}E_{x}+\alpha_{zy}E_{y}+\alpha_{zz}E_{z},
\end{array}\label{eq-comps-p-geral}
\end{equation}
which can be expressed in matrix notation as
\begin{equation}
\left[\begin{array}{c}
p_{x}\\
p_{y}\\
p_{z}
\end{array}\right]=\left[\begin{array}{ccc}
\alpha_{xx} & \alpha_{xy} & \alpha_{xz}\\
\alpha_{yx} & \alpha_{yy} & \alpha_{yz}\\
\alpha_{zx} & \alpha_{zy} & \alpha_{zz}
\end{array}\right]\left[\begin{array}{c}
E_{x}\\
E_{y}\\
E_{z}
\end{array}\right]. \label{eq-p-beta-geral-E}
\end{equation}
Note that, in this case, we have
\begin{equation}
\boldsymbol{\alpha}=\left[\begin{array}{ccc}
\alpha_{xx} & \alpha_{xy} & \alpha_{xz}\\
\alpha_{yx} & \alpha_{yy} & \alpha_{yz}\\
\alpha_{zx} & \alpha_{zy} & \alpha_{zz}
\end{array}\right] \label{eq-beta-geral}
\end{equation}
where all the coefficients  $\alpha_{ij}$ (with $i,j = x,y,z$) of this matrix, can be non nulls.
%
Despite this, the polarizability tensor has, in general, at most six independent components.
This is a consequence of the fact that the polarizability tensor is a symmetric tensor, which means that its elements $\alpha_{ij}$ have the property
\begin{equation}
\alpha_{ij} = \alpha_{ji}, \label{eq-symmetry}
\end{equation}
as showed in Eqs. \eqref{eq-beta-anisotr-rot} and \eqref{eq-symmetry-tilde}.


\subsection{Matrix representation of the polarizability tensor for a $\text{CO}_{2}$ molecule in a rotated coordinate system}
\label{rotated-system}

\begin{figure}[h]
\centering
\epsfig{file=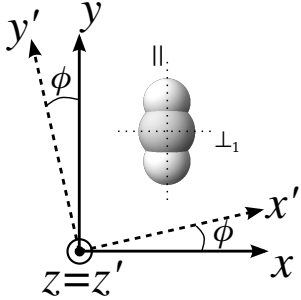,width=0.45 \linewidth}
\caption{\footnotesize{
A visualization of the system $x^\prime y^\prime z^\prime$ rotated with respect to $xyz$. 
Here the axis $z^\prime=z$ appears perpendicular to the paper.				
}}
\label{O-P-xyz-primo-com-CO2}
\end{figure}

In Sec. \ref{sec-scalar}, we presented the distance $d$ as an example of a scalar, 
in the sense that it is an invariant under the rotation of the coordinate system. 
In Sec. \ref{sec-vector}, we defined a vector as an object whose components transform,
under a rotation of the coordinate system, in the same manner
that the coordinates of a point in space.
In this section, we return in considering the molecule as oriented 
in Sec. \ref{representing-tensor} and \ref{visulization},
and discuss how the representation of the tensor $\overleftrightarrow{\alpha}$,
given by the matrix $\boldsymbol{\alpha}$ in Eq. \eqref{eq-p-E-matrix-anisotr}, 
transforms under a rotation of the coordinate system, as illustrated 
in Fig. \ref{O-P-xyz-primo-com-CO2}.
We remark that, in the present case, 
the $\text{CO}_{2}$ molecule stays put in space, whereas the coordinate system is rotated, so that
the tensor  $\overleftrightarrow{\alpha}$ itself has not been changed,
but its description in  $xyz$ is different from that in $x^\prime y^\prime z^\prime$.

\begin{figure}[h]
	\epsfig{file=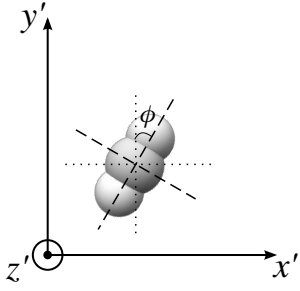, width=0.45 \linewidth}
	\caption{\footnotesize{
			Orientation of the $\text{CO}_{2}$ molecule, from the point of view of the system $x^\prime y^\prime z^\prime$.
	}}
	\label{fig:molecula-rotacao-para-sist-primo}
\end{figure}

Let us remember that, for the coordinate transformation in Fig. \ref{O-P-xyz-primo-com-CO2},
the relation between the coordinates $(x^\prime,y^\prime,z^\prime$) and
$(x,y,z)$ is given by the matrix $\mathbf{R}_0$ [see Eqs. \eqref{eq-x-primo-x-particular}
and \eqref{eq-R0-def}].
A shortcut to discover how is the matrix representation ${\boldsymbol{\alpha}^\prime}$, 
of the tensor $\overleftrightarrow{\alpha}$ in the system
$x^\prime y^\prime z^\prime$,
requires to note that, from the point of view of the system $x^\prime y^\prime z^\prime$, the $\text{CO}_{2}$ molecule 
seems as illustrated in Fig. \ref{fig:molecula-rotacao-para-sist-primo}.
Moreover, note that, the non-rotated $\text{CO}_{2}$ molecule seems to the system $x^\prime y^\prime z^\prime$ in the same way that a rotated $\text{CO}_{2}$ molecule seems to the system $xyz$, as shown in 
Fig. \ref{molecula-rotacao}.
Thus, in a similar reasoning used to find Eq. \eqref{eq-beta-anisotr-rot}, we can also obtain that 
${\boldsymbol{\alpha}^\prime}$, the matrix representation of the 
tensor $\overleftrightarrow{\alpha}$ in the system $x^\prime y^\prime z^\prime$, is given by
\begin{equation}
{\boldsymbol{\alpha}^\prime}=\left[\begin{array}{ccc}
		\alpha_{\parallel}\sin^{2}\phi+\alpha_{\perp}\cos^{2}\phi & (\alpha_{\parallel}-\alpha_{\perp})\sin\phi\cos\phi & 0\\
		(\alpha_{\parallel}-\alpha_{\perp})\sin\phi\cos\phi & \alpha_{\parallel}\cos^{2}\phi+\alpha_{\perp}\sin^{2}\phi & 0\\
		0 & 0 & \alpha_{\perp}
	\end{array}\right]. 
\label{eq-alpha-primo}
\end{equation}
It is direct to verify that
\begin{equation}
	{\boldsymbol{\alpha}^\prime}=\mathbf{R}^{(0)}\boldsymbol{\alpha}\mathbf{R}^{(0)T},
	\label{eq-alpha-primo-R0}
\end{equation}
which in tensor notation is written as ${\alpha}_{ij}^\prime=
\sum_{r=1}^3\sum_{s=1}^3 {R}^{(0)}_{ir}\alpha_{rs}{R}^{(0)T}_{sj}$,
which results
\begin{equation}
	{\alpha}_{ij}^\prime=\sum_{r=1}^3\sum_{s=1}^3{R}^{(0)}_{ir}R^{(0)}_{js}\alpha_{rs}.
	\label{eq-alpha-primo-R0-indices}
\end{equation}
Now, let us remember that the components of a vector transform as 
given in Eq. \eqref{eq-v-primo-v-index-R0}. 
Thus, we can write,
considering the components of two vectors $\textbf{v}$ and $\textbf{w}$,
\begin{equation}
	v_{i}^\prime w_{j}^\prime=\sum_{r=1}^3\sum_{s=1}^3{R}^{(0)}_{ir}R^{(0)}_{js}v_{r}w_{s}.
	\label{eq-vi-primo-vj-primo-R0-indices}
\end{equation}
Comparing Eqs. \eqref{eq-alpha-primo-R0-indices} and \eqref{eq-vi-primo-vj-primo-R0-indices},
we see that the quantities ${\alpha}_{ij}$ transform into ${\alpha}_{ij}^\prime$, under
the coordinate transformation \eqref{eq-x-primo-x-index-0}, like the product
of components of two vectors \cite{Landau-Lifshitz-vol-2}.

For a general rotation, as illustrated in Fig. \ref{rotacao-geral} and given by
Eqs. \eqref{eq-x-primo-x} and \eqref{eq-x-primo-x-index}, 
one has that Eqs. \eqref{eq-alpha-primo-R0} and \eqref{eq-alpha-primo-R0-indices},
are generalized, respectively, to
\begin{eqnarray}
	{\boldsymbol{\alpha}^\prime}&=&\mathbf{R}\boldsymbol{\alpha}\mathbf{R}^{T},\\
	{\alpha}_{ij}^\prime&=&\sum_{r=1}^3\sum_{s=1}^3{R}_{ir}R_{js}\alpha_{rs},
	\label{eq-alpha-primo-R-indices}
\end{eqnarray}
Thus, knowing the representation $\boldsymbol{\alpha}$ of the polarizability tensor 
relative to an arbitrarily set of axes,
we can know its representation ${\boldsymbol{\alpha}^\prime}$ in any other rotated system.
Adapting to this case the words of Feynman \cite{Feynman-Lectures-vol-2}, the polarizability of the $\text{CO}_{2}$ molecule
\begin{quote}
\textit{``...  is described completely by giving the components of the polarization tensor $\alpha_{ij}$ with respect to any arbitrarily chosen set of axes.''
}
\end{quote} 
And, just as we can associate a position vector $\textbf{r}=(x,y,z)$, or velocity $\textbf{v}=(v_x,v_y,v_z)$, with 
the molecule, so that the components of these vectors change in a certain definite way if we change the coordinate system [see Eq. \eqref{eq-v-primo-v-index}], so to the molecule we can also
\begin{quote}
	\textit{``...  associate its polarization tensor $\alpha_{ij}$, whose nine components will transform in a certain definite way if the coordinate system is changed.''
	}
\end{quote} 

Considering again a general rotation, given in
Eqs. \eqref{eq-x-primo-x} and \eqref{eq-x-primo-x-index}, Eq. \eqref{eq-vi-primo-vj-primo-R0-indices} is generalized to
\begin{equation}
	v_{i}^\prime w_{j}^\prime=\sum_{r=1}^3\sum_{s=1}^3{R}_{ir}R_{js}v_{r}w_{s}.
	\label{eq-vi-primo-vj-primo-R-indices}
\end{equation}
Comparing Eqs. \eqref{eq-alpha-primo-R-indices} and \eqref{eq-vi-primo-vj-primo-R-indices},
we see that the quantities ${\alpha}_{ij}$ transform into ${\alpha}_{ij}^\prime$, under
the coordinate transformation \eqref{eq-x-primo-x-index}, like the product
of components of two vectors \cite{Landau-Lifshitz-vol-2}.
In other words, a general second rank tensor is a set of nine quantities (in the present three-dimensional context)
which, under rotations of the coordinate system, transform like the products of the components
of two vectors \cite{Landau-Lifshitz-vol-2}.
%

Considering two vectors $\textbf{v}$ and $\textbf{w}$,
a second rank tensor can be built by means of the tensor product
$\textbf{v}\otimes\textbf{w}$, 
defined so that their components are \textit{``the products of the
	components of the two vectors of the product''} \cite{Cohen-QM-vol-1}. Thus,
\begin{equation}
	(\textbf{v}\otimes\textbf{w})_{ij}=v_i w_j	
	\label{eq:tensor-def}
\end{equation}
To illustrate the use of tensor products, let us use them to rewrite
the polarizability tensor in Eq. \eqref{eq:alpha-matrix-anisotr}, for
a $\text{CO}_{2}$ molecule shown in Fig. \ref{fig:molecula-eixo-y}.
From the definition in Eq. \eqref{eq:tensor-def}, we have:
\begin{equation}
	\hat{\textbf{x}}\otimes\hat{\textbf{x}}=\left[\begin{array}{ccc}
		1 & 0 & 0\\
		0 & 0 & 0\\
		0 & 0 & 0
	\end{array}\right],
\label{eq:x-otimes-x}
\end{equation}
\begin{equation}
	\hat{\textbf{y}}\otimes\hat{\textbf{y}}=\left[\begin{array}{ccc}
		0 & 0 & 0\\
		0 & 1 & 0\\
		0 & 0 & 0
	\end{array}\right], 
	\label{eq:y-otimes-y}
\end{equation}
\begin{equation}
	\hat{\textbf{z}}\otimes\hat{\textbf{z}}=\left[\begin{array}{ccc}
		0 & 0 & 0\\
		0 & 0 & 0\\
		0 & 0 & 1
	\end{array}\right]. 
	\label{eq:z-otimes-z}
\end{equation}
Then, we can write
\begin{equation}
	\boldsymbol{\alpha}=\alpha_\bot\hat{\textbf{x}}\otimes\hat{\textbf{x}}+\alpha_\parallel
	\hat{\textbf{y}}\otimes\hat{\textbf{y}}+\alpha_\bot\hat{\textbf{z}}\otimes\hat{\textbf{z}}. 
\end{equation}

\subsection{A general anisotropic particle}
\label{sec:general-anisotropic}

A general anisotropic object is one whose diagonal representation 
of its polarizability tensor is given by
\begin{equation}
	\boldsymbol{\alpha}=\left[\begin{array}{ccc}
		\alpha_a & 0 & 0\\
		0 & \alpha_b & 0\\
		0 & 0 & \alpha_c
	\end{array}\right]. \label{eq:alpha-geral-anisotr}
\end{equation}
where $\alpha_a\neq\alpha_b\neq\alpha_c$.
Let us consider the principal axis $a$, $b$, and $c$ (perpendicular to each other)
of this anisotropic object parallel to $x$, $y$, and $z$, respectively.
Under the action of an external electric field $\textbf{E}$,
we have the induced dipole $\textbf{p}$ given by $\textbf{p}={\textbf{p}}_{x}+{\textbf{p}}_{y}+{\textbf{p}}_{z}$,
where:
\begin{eqnarray}
	{\textbf{p}}_{x}&=&\alpha_{a}\textbf{E}_{x},
	\label{pa}\\
	{\textbf{p}}_{y}&=&\alpha_{b}\textbf{E}_{y},
	\label{pb}\\
	{\textbf{p}}_{z}&=&\alpha_{c}\textbf{E}_{z}.
	\label{pc}
\end{eqnarray}
Using Eqs. \eqref{pa}, \eqref{pb}, and \eqref{pc}, in
\begin{equation}
	E^{2}={E}_{a}^2+{E}_{b}^2+{E}_{c}^2,
	\label{E-aniso}
\end{equation}
we have
\begin{equation}
	\frac{p_{x}^{2}}{E^{2}\alpha_{a}^{2}}+\frac{p_{y}^{2}}{E^{2}\alpha_{b}^{2}}+	\frac{p_{z}^{2}}{E^{2}\alpha_{c}^{2}}=1.
	\label{pre-elipsoide-aniso}
\end{equation}
Choosing $\alpha_{a}<\alpha_{c}<\alpha_{b}$, and defining
\begin{eqnarray}
	p_{a\text{(min)}}^{2}=E^{2}\alpha_{a}^{2},
	\label{p-a-min-def}
	\\
	p_{b{\text{(max)}}}^{2}=E^{2}\alpha_{b}^{2},
	\label{p-b-int-def}
		\\
	p_{c{\text{(int)}}}^{2}=E^{2}\alpha_{c}^{2},
	\label{p-c-max-def}
\end{eqnarray}
we can rewrite \eqref{pre-elipsoide-aniso} as
\begin{equation}
	\frac{p_{x}^{2}}{p_{a{\text{(min)}}}^{2}}+\frac{p_{y}^{2}}{p_{b{\text{(max)}}}^{2}}+	\frac{p_{z}^{2}}{p_{c{\text{(int)}}}^{2}}=1.
	\label{eq:elipsoide-aniso}
\end{equation}
Note that the minimum magnitude of $\textbf{p}$, $	p_{a\text{(min)}}$, 
occurs when the field $\textbf{E}$ is parallel to the $a$-axis, whereas the maximum,
$p_{b{\text{(max)}}}$, when the field is parallel to the $b$-axis. A certain intermediate
value, $p_{c{\text{(int)}}}$, occurs when the field $\textbf{E}$ is parallel to the $c$-axis.
%

\section{Summary}
\label{sec:summary}
\begin{figure}[h]
	\epsfig{file=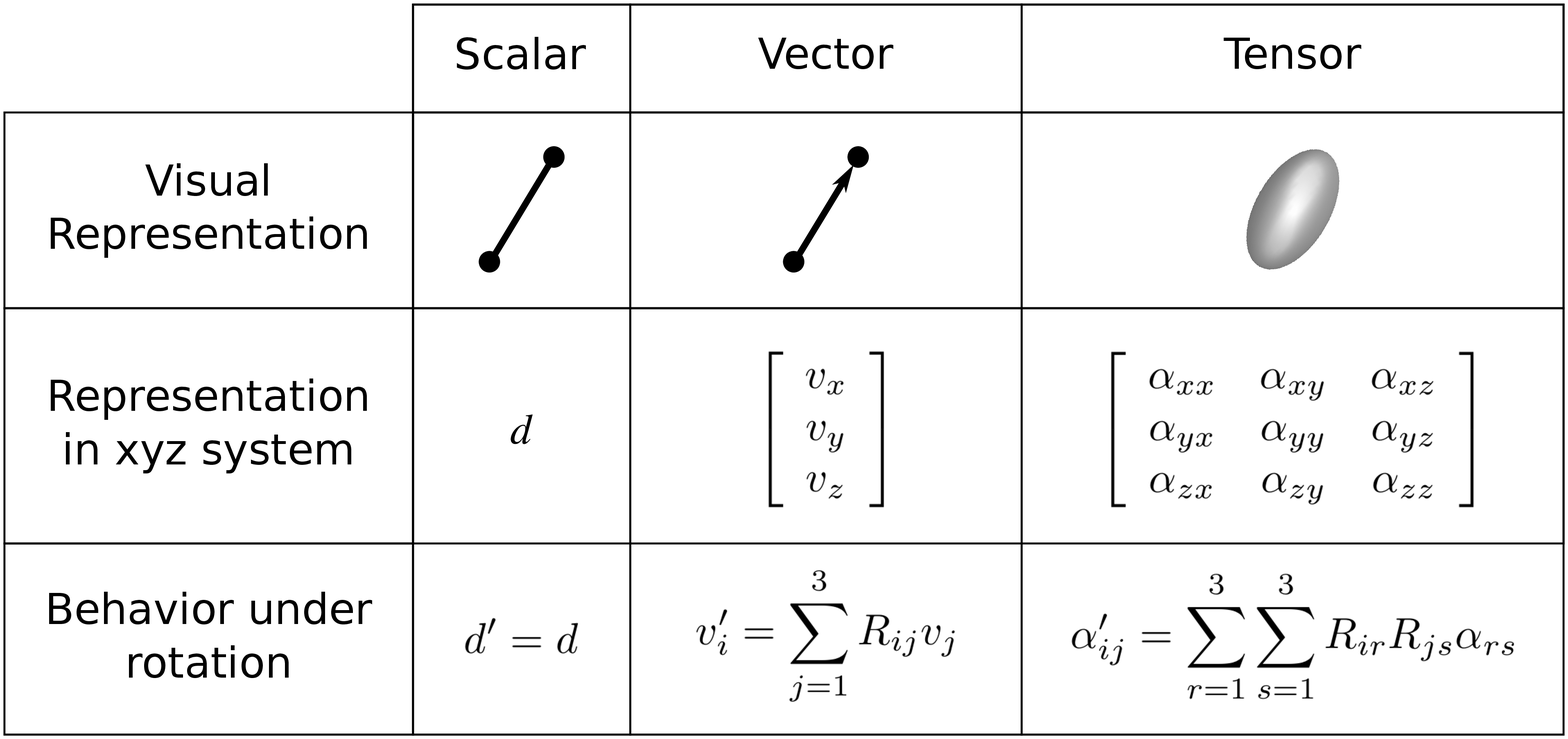, width=0.90 \linewidth}
	\caption{\footnotesize{
			Summary of the main ideas discussed in this paper.
			Each column exhibits information about
			a given mathematical object: a scalar (column 1), a vector (column 2),
			and a second-rank tensor (column 3).
			In line 1, we show, for each mathematical object, a certain visual representation. 
			In line 2, it is shown the numerical representations in a 
			coordinate system $xyz$. 
			In line 3, we show the numerical representations in a 
			coordinate system $x^\prime y^\prime z^\prime$, rotated in relation to
			$xyz$ according to Eq. \eqref{eq-x-primo-x-index}.
	}}
	\label{fig:tabela}
\end{figure}

Inspired by Ref. \cite{Fleisch-StudentsGuide-2011}, we organize, in the table
shown in Fig. \ref{fig:tabela}, some basic ideas about a scalar, vector,
and second-rank tensor, thus summarizing the main ideas discussed in this paper.


\section{Final remarks} 
\label{sec-final}

We proposed an introduction to the notion of tensors, 
inspired by the Feynman didactic approach found in Ref. \cite{Feynman-Lectures-vol-2},
but with some variations.
Instead of crystals, as the base models, we considered a single ground-state atom 
(Sec. \ref{isotropic}) and a  $\text{CO}_{2}$ molecule (Sec. \ref{polarizability-anisotropic-particle}), and introduced a visual representation of tensors based on the ideas of the Lamé stress ellipsoid
(Sec. \ref{visulization}), rather than the energy ellipsoid.
%

To deal with a single atom or molecule, just the dipole moment $\textbf{p}$ is necessary
[as discussed in Eqs. \eqref{p-beta-E} and \eqref{p-beta-E-tensor}],
whereas to deal with crystals, 
the dipole moment per unit volume $(\textbf{P})$ is considered in Ref. \cite{Feynman-Lectures-vol-2}.
The perception of the isotropic polarizability of an
atom with a spherically symmetric electron cloud, as discussed using Fig.
\ref{atomo}, is more direct than that, for instance, of a cubic crystal.
The visualization of the polarizabilities along the principal axes
of a single $\text{CO}_{2}$ molecule, exploring its symmetries as shown in Fig. \ref{fig:molecula-eixos-principais}, is more straightforward than dealing with a crystal.
The visual representation of the $\text{CO}_{2}$ polarizability tensor was done here 
by the construction of the correspondent Lamé ellipsoid,
as shown in Eqs. \eqref{eq:ellipsoid-passo-0}-\eqref{elipsoide-3d},
which did not require differential calculus, just introductory vector algebra.
In counterpart, to deal with the energy ellipsoid, as discussed in Ref. \cite{Feynman-Lectures-vol-2}, 
it is required some notion of differential and integral calculus, and also ideas on the energy per unit volume required to polarize a crystal.

In conclusion, the introduction to tensors presented here
requires less mathematical tools and physical concepts than the original Feynman approach.
Thus, it can be used by students still in earlier levels, and helping them 
to follow the original Feynman approach \cite{Feynman-Lectures-vol-2}. 
\appendix

\section{A general form for $\textbf{R}$}
\label{ap:R}
For a general rotated Cartesian coordinate system $x^\prime y^\prime z^\prime$ (as illustrated in Fig. \ref{rotacao-geral}), the relation between the coordinates $(x^\prime,y^\prime,z^\prime$) and
$(x,y,z)$ takes the form in Eq. \eqref{eq-x-primo-x}
where the square matrix $\mathbf{R}$
is orthogonal.
Although the explicit form of $\mathbf{R}$ is not necessary to follow the reasoning
through the main text of this article, 
we exhibit, for informational purposes, the general aspect of this matrix in terms of Euler angles
(for more details about Euler angles, see, for instance, Refs. \cite{Landau1-Lifshitz-vol-1,Goldstein-CM,Sakurai-QM-1994, Ballentine-QM-2014}).
Denoting the Euler angles by $(\phi,\theta,\psi)$, 
according to the convention usually adopted in quantum mechanics 
\cite{Sakurai-QM-1994, Ballentine-QM-2014}, we have
$x_i^{\prime}=\sum_j R_{ij}x_j$,
where $R_{ij}$ are the elements of the Euler rotation matrix $R(\phi,\theta,\psi)$,
given by
\begin{eqnarray}
R_{11}&=&\cos(\theta)\cos(\psi)\cos(\phi)-\sin(\psi)\sin(\phi),\nonumber\\
R_{12}&=&\cos(\theta)\cos(\psi)\sin(\phi)+\cos(\phi)\sin(\psi),\nonumber\\
R_{13}&=&-\sin(\theta )\cos(\psi),\nonumber\\
R_{21}&=&-\cos(\psi)\sin(\phi)-\cos(\theta)\sin(\psi)\cos(\phi),\nonumber\\
R_{22}&=&\cos(\psi)\cos(\phi)-\cos(\theta)\sin(\psi)\sin(\phi),\nonumber\\
R_{23}&=&\sin(\theta)\sin(\psi),\nonumber\\
R_{31}&=&\sin(\theta)\cos(\phi),\nonumber\\
R_{32}&=&\sin(\theta)\sin(\phi),\nonumber\\
R_{33}&=&\cos(\theta).\nonumber
\end{eqnarray}	
Note that, when considering $\theta=0$ and $\psi=0$, we obtain
%
%
%
\begin{eqnarray}
R_{11}&=&\cos(\phi),\nonumber\\
R_{12}&=&\sin(\phi),\nonumber\\
R_{13}&=&0,\nonumber\\
R_{21}&=&-\sin(\phi),\nonumber\\
R_{22}&=&\cos(\phi),\nonumber\\
R_{23}&=&0,\nonumber\\
R_{31}&=&0,\nonumber\\
R_{32}&=&0,\nonumber\\
R_{33}&=&1,\nonumber
\end{eqnarray}
and we recover the matrix $\textbf{R}^{(0)}$ [given in Eq. \eqref{eq-R0-def}], showing that $\textbf{R}$ is a generalization of this matrix.

\begin{acknowledgments}

%
L.Q. and E.C.M.N. were supported by the Coordenação de Aperfeiçoamento de Pessoal de Nível Superior - Brasil (CAPES), Finance Code 001.
%
%
\end{acknowledgments}
%



%

\end{document}